\documentclass[11pt,a4paper]{article}
\usepackage{jheppub}
\usepackage{graphicx}
\usepackage{lipsum}
\usepackage{slashed}
\usepackage{mathtools}
\usepackage{bm}
\usepackage[all]{hypcap}
\usepackage{tikz}
\usepackage[compat=1.1.0]{tikz-feynman}
\usepackage{bbm}
 \usepackage[table,xcdraw,dvipsnames]{xcolor}
\usepackage{soul}
\usepackage{mathtools}
\usepackage{colortbl}
\usepackage{booktabs}
\usepackage{subcaption}
\usepackage[capitalize]{cleveref}
\usepackage[font=small,labelfont=bf]{caption}

\usepackage{parskip}
\newcommand{\NDW}{{N_\textrm{DW}}}
\newcommand{\A}{{\cal A}}

\newcommand{\Z}{{\mathbb Z}}

\newcommand{\PQ}{{\rm PQ}}

\newcommand{\alphaem}{\alpha_{\textrm{em}}}

\definecolor{blus}{cmyk}{1,1,0,0.6}
\definecolor{green}{cmyk}{0.92,0,0.59,0.25}
\definecolor{red}{cmyk}{0,1,1,0.55}
\definecolor{nicepurple}{RGB}{193,67,213}

\hypersetup{colorlinks,bookmarksopen,bookmarksnumbered,linkcolor=green,pdfstartview=FitH,urlcolor=green,citecolor=green}

\newcommand{\beq}{\begin{equation}}
\newcommand{\eeq}{\end{equation}}
\newcommand{\bea}{\begin{eqnarray}}
\newcommand{\eea}{\end{eqnarray}}

\def\be{\begin{equation}}
\def\ee{\end{equation}}

\title{
Axion Quality from Exact Proton Stability
}

\author[a,b]{Joe Davighi,}\emailAdd{joseph.davighi@cern.ch}
\author[c]{Admir Greljo,}\emailAdd{admir.greljo@unibas.ch}
\author[c]{and Xavier Ponce D\'iaz}\emailAdd{xavier.poncediaz@unibas.ch}

\affiliation[a]{Theoretical Physics Department, CERN, 1211 Geneva 23, Switzerland}
\affiliation[b]{DAMTP, University of Cambridge, Wilberforce Road, Cambridge CB3 0WA}
\affiliation[c]{Department of Physics, University of Basel, Klingelbergstrasse 82, CH 4056 Basel, Switzerland}

\abstract{
A discrete, anomaly-free $\mathbb Z_9$ gauge symmetry may persist to the deep infrared, exactly forbidding proton decay. This $\mathbb Z_9$ can emerge from Higgsing a lepton-flavour non-universal $U(1)_X$ that generates realistic neutrino masses and mixings through a high-scale seesaw and predicts an additional light Goldstone. Adding an anomaly-free chiral heavy-quark sector turns this mode into the QCD axion. The same selection rules that ban proton decay also forbid PQ-violating operators below a suitably high dimension, $d_{\rm PQ}\gtrsim10$, including operators involving the heavy quarks that radiatively match onto the scalar potential. We construct explicit models viable when PQ breaking occurs before or after inflation; in the latter case, avoiding stable heavy relics and eliminating the string--wall network tightly constrain the heavy-quark spectrum. We identify restricted regions of parameter space where thermal leptogenesis remains viable and, for $d_{\rm PQ}\geq13$, can coexist with a high-quality axion constituting all of the dark matter.
}

\begin{document}

\maketitle

\newpage

\section{Introduction}

We do not know, experimentally, the global structure of the Standard Model (SM) gauge group. While it is often assumed to be connected,\footnote{Even if we assume it is connected, a four-fold ambiguity remains~\cite{Tong:2017oea} that could be disentangled, for instance, by the discovery of fractionally charged particles~\cite{Davighi:2019rcd,Koren:2024xof,Alonso:2024pmq,Koren:2025utp}.} there may in fact be a remnant discrete gauge symmetry arising {\em e.g.}~from an incompletely Higgsed continuous symmetry in the ultraviolet,  that persists to the deep infrared. This is an interesting possibility, phenomenologically: a discrete gauge symmetry would give sharp selection rules on the low-energy EFT without any accompanying light degrees of freedom or constraint on the EFT cut-off.

In fact, the persistent non-observation of proton decay may be interpreted as a hint of such an exact selection rule (rather than hinting merely that baryon-number violation occurs but at a very high scale). It has been shown that gauging an accidental $\Gamma \cong \mathbb{Z}_9$ symmetry of the SM, under which all quarks carry unit charge while leptons and the Higgs are neutral, enforces the exact selection rule
\begin{equation}
\Delta B = 0 \,\, (\textrm{mod}\,3)\, ,
\end{equation}
thereby forbidding all $\Delta B = 1$ or $2$ transitions {\em to all orders} in the EFT~\cite{Babu:2003qh,Davighi:2022qgb, Koren:2022bam}. 

Interpreting proton stability as a low energy remnant of far UV dynamics has also proven useful in addressing other empirical observations that point to physics beyond the Standard Model (BSM). In particular, the proton-stabilising symmetry $\Gamma$ can arise naturally from Higgsing an anomaly-free $U(1)_X$ whose breaking generates Majorana masses for right-handed neutrinos near the GUT scale. As shown in~\cite{Greljo:2025suh}, the same minimal setup can simultaneously accommodate neutrino masses and mixings, generate the baryon asymmetry through thermal leptogenesis, and provide a dark-matter candidate in the form of the associated Nambu-Goldstone boson (NGB), the majoron, without modifying the field content or basic structure originally introduced to ensure exact proton stability~\cite{Davighi:2022qgb}.

An intriguing feature of this construction is that the same symmetry-breaking sector already contains a light NGB. This raises the question of whether the majoron can, in addition, play the role of the QCD axion~\cite{Weinberg:1977ma,Wilczek:1977pj}, thereby addressing another outstanding puzzle of the SM: the strong CP problem. By introducing heavy coloured fermions, vector-like under the SM but chiral under $U(1)_X$, the Goldstone can acquire the required mixed anomaly with QCD, in close analogy with Kim~\cite{Kim:1979if},
Shifman, Vainshtein and Zakharov~\cite{Shifman:1979if} (KSVZ) axion models.

This possibility immediately raises the axion quality problem~\cite{Georgi:1981pu,Dine:1986bg,Barr:1992qq,Holman:1992us,Kamionkowski:1992mf,Ghigna:1992iv}. Since the Peccei--Quinn (PQ) symmetry~\cite{Peccei:1977hh,Peccei:1977ur} is global and accidental, generic UV effects, and in particular those associated with quantum gravity, are expected to generate PQ-violating operators. Unless sufficiently suppressed, these contributions modify the axion potential and displace its minimum away from the CP-conserving point, thereby spoiling the PQ solution to the strong CP problem. The remarkable smallness of $\bar\theta$ therefore requires explicit PQ breaking to be absent, or strongly suppressed, up to rather high operator dimension.

A well-known strategy to address this problem is to realise $U(1)_\PQ$ as an accidental global symmetry enforced by an exact gauge symmetry. In this way, gauge invariance can forbid the lowest-dimensional PQ-breaking operators, postponing explicit breaking to sufficiently high order. This idea has been explored extensively in models based on additional gauged $U(1)$ symmetries~\cite{Barr:1992qq,Fukuda:2017ylt,Qiu:2023los,Albertus:2026fbe,Babu:2026yqp}, including constructions in which the protecting gauge symmetry is itself associated with flavour~\cite{Babu:1992cu,Cheung:2010hk}, $B-L$~\cite{Ibe:2018hir}, or baryon number~\cite{Duerr:2017amf}. Other possibilities include protection by discrete gauge symmetries~\cite{Chun:1992bn,Babu:2002ic,Dias:2002gg}, non-Abelian gauge symmetries~\cite{DiLuzio:2017tjx,Ardu:2020qmo,Darme:2021cxx}, extra-dimensional theories~\cite{Choi:2003wr,Cox:2019rro,Craig:2024dnl,Craig:2026dcb,FernandezNavarro:2026bed}, or composite axion constructions in which PQ emerges accidentally from the underlying gauge dynamics~\cite{Gavela:2018paw, Agrawal:2025mke}. In this paper, we show that the same anomaly-free $U(1)_X$ whose breaking leaves the proton-stabilising $\mathbb{Z}_9$ also controls the accidental PQ symmetry and the dimension of its leading explicit breaking, and can offer an excellent solution to the axion quality problem.\footnote{Interestingly, a related connection between axion quality and proton stability was explored early on in supersymmetric models~\cite{Chun:1992bn}, where discrete gauge symmetries were used to obtain an automatic Dine--Fischler--Srednicki--Zhitnitsky-like (DFSZ) PQ symmetry~\cite{Zhitnitsky:1980tq,Dine:1981rt} while suppressing dangerous proton-decay operators and accommodating neutrino masses. The mechanism is, however, distinct from the exact $\Delta B=0~(\mathrm{mod}\,3)$ selection rule considered here.}

We construct explicit realisations of this idea and explore their phenomenological and cosmological implications. In \cref{sec:discrete}, we review the proton-stabilising $U(1)_X$ framework and its connection to realistic neutrino masses, and identify the conditions under which the same setup can support thermal leptogenesis. In \cref{sec:2scalars}, we show that, upon introducing heavy coloured fermions vector-like under the SM but chiral under $U(1)_X$, the associated NGB can be promoted to a KSVZ-like QCD axion. Crucially, the same $U(1)_X$ charge assignments that determine the proton-stabilising remnant and the neutrino sector also fix the leading gauge-invariant PQ-breaking operator. The axion quality can therefore be quantified directly in terms of the underlying charge structure, without introducing an additional symmetry solely for this purpose. We furthermore extend the quality analysis beyond the scalar potential by considering PQ-violating operators involving the heavy fermions, which can shift the axion potential via loop effects~\cite{Bonnefoy:2022vop}. We then study the resulting phenomenology in the pre-inflationary scenario, identifying regions in which high axion quality remains compatible with the relevant axion, cosmological and leptogenesis constraints.

In \cref{sec:post_inflationary}, we turn to the more demanding case in which the symmetry is broken after inflation. In this scenario, the formation of topological defects and the thermal production of the heavy coloured states introduce the additional domain-wall and stable relic problems~\cite{Sikivie:1982qv,Vilenkin:1982ks,Kibble:1976sj,Lu:2023ayc}. We show that these requirements can be addressed simultaneously by constructing anomaly-free heavy-fermion spectra that allow the exotic states to decay sufficiently rapidly while preserving the required axion quality. We study the resulting string--domain-wall dynamics and construct explicit benchmark models in which high axion quality, exact proton stability and realistic neutrino masses coexist with a viable post-inflationary cosmological history. Remarkably, regions of parameter space remain in which the same setup can also account for the baryon asymmetry through thermal leptogenesis and for the dark-matter abundance through the QCD axion. We discuss future prospects at various kinds of experiment, before concluding in \cref{sec:concl}.

\section{Discrete gauge symmetries and proton stability }
\label{sec:discrete}

Consider gauging an anomaly-free $U(1)_X$ subgroup of the SM's accidental symmetry:
\begin{equation} \label{eq:U1X}
    X = 3mB - 9 \sum_{i=1}^3 r_i L_i, \qquad \sum_i r_i = m\, .
\end{equation}
The overall factor of 3 is just so that all charges are integral (quarks have $B=1/3$), and $\sum_i r_i = m$ is required by anomaly cancellation. If
\begin{equation}
    \gcd(m,9) = 1, 
\end{equation}
{\em i.e.} provided $m$ is not divisible by 3, then we can define a $\Z_9$ subgroup of $X$ acting as 
\begin{equation}
    q \mapsto e^{2\pi i/9}q, \qquad \ell,\, H \mapsto \ell,\, H\,,
\end{equation}
where $q$ ($\ell$) denotes any SM quark (lepton) field, and $H$ is the SM Higgs. Note that the lepton flavour universal points $r_{e}=r_\mu=r_\tau$ imply $m \in 3\Z$ and so are incompatible. Thus, lepton flavour {\em non}-universality, and in particular there being $3n$ generations of SM fermion, is a necessary condition to find this $\Z_9$ symmetry.

We suppose $U(1)_X$ is spontaneously broken at a high scale by scalar fields $\phi_i$ whose charges $X_i$ satisfy
\begin{equation} \label{eq:Xgcd}
    \gcd(X_i) \in 9 \Z\, .
\end{equation}
A generic condensate $\langle \phi_i \rangle$ would then break
\begin{equation}
    \langle \phi_i \rangle : U(1)_{X} \to \Gamma \supset \Z_9 \,,
\end{equation}
where the discrete gauge symmetry contains the $\Z_9$ subgroup acting only on quarks, which have charge $1\pmod9$. This discrete gauge symmetry guarantees the exact selection rule
\begin{equation}
    \Delta B = 0 \pmod3
\end{equation}
is satisfied by all amplitudes~\cite{Davighi:2022qgb,Koren:2022bam}. This would predict that protons are exactly stable and forbid neutron-antineutron oscillations. It allows, however, baryon-number violation induced by the electroweak sphaleron process, since it violates $B$ in units of 3.

\subsection{Connection to neutrino masses}
\label{sec:neutrino_texture}

An option for the scalar charges consistent with~\eqref{eq:Xgcd} is for each $X_i$ to be the sum of two lepton charges, 
\begin{equation}\label{eq:neutrino_mass_condition}
    X_i \in \{9(r_j + r_k)\}
\end{equation}
where $j, k \in e,\mu,\tau$ (the case $j=k$ is not excluded). With such charges, the UV theory permits Yukawa interactions that give Majorana masses to RH neutrinos, in a flavor-dependent way, at the $U(1)_X$ breaking scale. Schematically,
\begin{equation}
    \mathcal{L} \supset \bar{N}_j^c N_k \phi_i \, .
\end{equation}
Generally speaking, as long as there are {\em at least two} such scalar fields, with distinct charges in the set~\eqref{eq:neutrino_mass_condition}, we can populate a realistic neutrino mass matrix~\cite{Davighi:2022qgb, Greljo:2025suh}.

For the rest of the paper, it will be useful to describe some explicit scenarios for the $r_i$ assignments that work well for neutrinos.
The most minimal scenarios that realise an acceptable PMNS matrix and neutrino mass texture, explored in~\cite{Davighi:2022qgb, Greljo:2025suh}, are obtained by setting two of the $r_i$ to be the same:
\begin{equation} 
\label{eq:r2r3}
    r_i=r_j\equiv r_{D}, \qquad r_k = r_{S}\neq r_{D}\, ,
\end{equation} 
where $i,j,k$ is some permutation of $e,\mu,\tau$. Here `D' stands for `doublet', and `S' for `singlet'. For these scenarios, characterised by having lepton flavour universality between two generations, the two scalar charges can be chosen to generate respectively the $N_1N_{2,3}$ and $N_{2,3}N_{2,3}$ entries of the right-handed neutrino mass matrix, 
\begin{equation} \label{eq:r2equalsr3}
X_1=9(r_{S}+r_{D}) \, ,
\qquad
X_2=18r_{D} \, .
\end{equation}
The requirement that the scalar vacuum expectation values (VEVs) break $U(1)_X$ precisely to the proton-stabilising $\mathbb Z_9$ implies
\begin{equation} \label{eq:gcd_2flavour}
\gcd(r_{S}+r_{D},2r_{D})=1\, .
\end{equation}
Note that, if we have normalised the underlying charges such that $\gcd(r_{D},r_{S})=1$, then condition~\eqref{eq:gcd_2flavour} is equivalent to requiring $r_{D}+r_{S}$ is odd, {\em i.e.} $r_{D}$ and $r_{S}$ have opposite parity.

However, if $r_{D}$ and $r_{S}$ have the same parity, coprimality implies that they are both odd, and therefore $\gcd(r_{S}+r_{D},2r_{D})=2\,.$ In this case the scalar charges have an additional common factor of two and the breaking pattern is instead $U(1)_X\longrightarrow \mathbb Z_{18} \,$. The additional $\mathbb Z_2$ factor acts as $(-1)^{X}=(-1)^F$ on the field content of the theory: all fermions carry odd $X$ charge, whereas the scalar fields carry even charge. It therefore corresponds simply to fermion parity and does not impose additional selection rules on operators beyond those of the proton-stabilising $\mathbb Z_9$.

With this charge assignment, the texture for the RH neutrino mass matrix can be written as
\begin{equation}
\label{eq:text_1}
    M_R\sim \begin{pmatrix}
        0 & 
        \phi_1 & \phi_1 \\
        \phi_1 & \phi_2 & \phi_2 \\
        \phi_1 & \phi_2 & \phi_2 
    \end{pmatrix} \,,
\end{equation}
where Yukawas have been omitted.\footnote{Note that, at the level of this discussion, models with $r_{S}=r_\mu$ or $r_{S}=r_\mu$ can be recovered by the use of permutation matrices, see Ref.~\cite{Greljo:2025suh}.} In this symmetry, we note that the lightest RH neutrino mass is parametrically lighter. In particular, with a mild hierarchy $v_2 >v_1$, and assuming generic Yukawa couplings $\sim 1$ the mass of the light RH neutrino is
\begin{equation}
    \label{eq:MN1_text_1}
    M_{N_1} \simeq \frac{v_1^2}{v_2} \, .
\end{equation}
This hierarchy in the spectrum will become important for our discussion on leptogenesis below.

Finally, the presence of zeros in both the Majorana and Dirac textures induces correlations among the low-energy parameters of the neutrino mass matrix. Most notably, \cref{eq:text_1} implies at tree level a relation between the mass of the lightest LH neutrino and the matrix element governing neutrinoless double-beta decay~\cite{Greljo:2025suh}. In particular, of the three possible charge assignments and two possible orderings, only three can be fitted within current neutrino oscillation data, cosmological measurements, and neutrinoless double-beta decay. Those models are: $e-$specific Inverted Ordering, and $\mu-$ and $\tau-$specific in Normal Ordering~\cite{Greljo:2025suh}. They make interesting predictions for the future neutrino experiments, as summarised in Fig.~2 of~\cite{Greljo:2025suh}.

\subsection{Guidance from leptogenesis}

This class of models already has the ingredients for an economical explanation of the baryon asymmetry of the Universe (BAU). The seesaw structure introduces new sources of CP violation through the complex neutrino Yukawa couplings and Majorana mass terms. The CP-violating decays of the heavy RH neutrinos can then generate a lepton asymmetry~\cite{Fukugita:1986hr}, provided that the RH-neutrino population develops a sufficient departure from thermal equilibrium. Electroweak sphaleron processes then partially convert this lepton asymmetry into a baryon asymmetry~\cite{Kuzmin:1985mm,Harvey:1990qw}. Together, lepton-number violation from the Majorana masses, CP violation in the RH-neutrino decays, and the departure from thermal equilibrium realize the three Sakharov conditions required for baryogenesis~\cite{Sakharov:1967dj}.

An additional relevant ingredient in the present class of models is the gauged $U(1)_X$, because gauge-mediated processes such as $N_iN_j \xleftrightarrow{~X_\mu~} f\bar f$ can keep the RH neutrinos close to thermal equilibrium~\cite{Racker:2008hp, Heeck:2016oda}. While this helps provide a thermal population of RH neutrinos at early times, if the gauge interaction remains efficient after the RH neutrinos become non-relativistic, it will delay their departure from equilibrium and can consequently suppress the efficiency of leptogenesis. Successful thermal leptogenesis therefore requires the $U(1)_X$ interactions to decouple sufficiently early, providing a direct connection between the symmetry-breaking scale and BAU generation. 

One can quantitatively estimate the conditions on the parameters of the model for it to realise thermal leptogenesis. The thermally-averaged cross-section for $NN \leftrightarrow f \bar{f}$ annihilations, assuming $M_{X}\gg \Gamma_{X}$ and $M_X \gtrsim 2 M_{N_1}$, is~\cite{Heeck:2016oda}
\begin{equation}
    \langle \sigma v\rangle \simeq \frac{g_X^4 X_\textrm{SM}^2\, X_{N_1}^2}{\pi}\frac{M_{N_1}T}{M_{X}^4}\, ,
\end{equation}
with $X_{N_1}=-9r_{S}$ the $U(1)_X$ charge of the lightest RH neutrino, after neglecting $v_1/v_2$ suppressed rotations and 
\begin{equation}
    X_\textrm{SM}^2 = \sum_f N_f X_f^2 = 18 m^2+\frac{243}{2} \sum_i r_i^2\, .
\end{equation}
The rate of annihilations relative to the Hubble rate is then approximately
\begin{equation}
\label{eq:gauge_decoupling}
   \left. \frac{\Gamma}{H} \right|_{M_{N_1}} \simeq 8.57 X_\textrm{SM}^2r_{S}^2 \left(\frac{g_X}{0.1}\right)^4\left(\frac{10^9\,\textrm{GeV}}{M_{N_1}}\right) \left(\frac{10 M_{N_1}}{M_{X}}\right)^4\, ,
\end{equation}
where we need to impose $\Gamma/H< 1$ for the decays to happen out of equilibrium. Note that in this approximation $g_X$ drops out from the mass of the $X$ boson. 

In addition, standard hierarchical thermal leptogenesis is subject to the Davidson--Ibarra (DI) bound~\cite{Davidson:2002qv}, which limits the maximal CP asymmetry that can be generated in $N_1$ decays. Reproducing the observed BAU therefore implies a lower bound on the RH-neutrino mass. Assuming a thermal initial population, and a hierarchical right-handed neutrino spectrum, one finds~\cite{Giudice:2003jh}
\begin{equation}
\label{eq:dav_ib}
M_{N_1}\gtrsim 4.9\times 10^8\,\textrm{GeV}\, .
\end{equation}
Requiring both \cref{eq:gauge_decoupling,eq:dav_ib} places constraints on the VEVs of the two scalar fields. For instance, a hierarchy between $v_2$ and $v_1$ is required to naturally decouple the $X$ interactions by the time $T\simeq M_{N_1}$. However, the DI bound prevents this hierarchy from becoming arbitrarily large, as can be seen from \cref{eq:MN1_text_1,eq:dav_ib}. The same constraints apply to the second texture, with the caveat that the Majorana mass in the texture remains a free parameter.

More precisely, for the texture in \cref{eq:text_1}, in the limit of $v_2\gg v_1$ the mass of the $X$ boson can be simplified to $M_{X}\simeq g_X |X_2| v_2$, and hence the out-of-equilibrium condition translates to an inequality on the ratio of scales:
\begin{equation} \label{eq:leptogen}
    \frac{v_1}{v_2} \lesssim 0.15 \left(\frac{X_2^2}{100}\right)^{1/3}\left(\frac{100}{X_\textrm{SM}^2 r_{S}^2}\right)^{1/6} \left(\frac{v_2}{10^{11}
    \, \textrm{GeV}}\right)^{1/6} \, .
\end{equation}
Modulo extremely large charge ratios, this typically means we need $v_1/v_2\lesssim \mathcal{O}(10^{-1})$, in other words, some modest scale hierarchy, for leptogenesis.

\section{From Majoron to Axion: Pre-inflationary models} \label{sec:2scalars}

We have seen in the previous Section that, by making the connection between the proton-stabilising $\Z_9$ symmetry and observed neutrino masses and mixings, we are led to consider at least two scalar fields that break the gauged $U(1)_X \to \Z_9$ at the high scale. Breaking a single gauged $U(1)_X$ with two complex scalars implies there is a light scalar Goldstone mode: a majoron. 

The phenomenology of this setup was studied in Ref.~\cite{Greljo:2025suh}, where the majoron provides a dark-matter candidate, while neutrino masses and the matter--antimatter asymmetry arise from the same high-scale seesaw sector predicting neutrino-mass textures compatible with oscillation data and minimal thermal leptogenesis, see also recent works~\cite{Akita:2026gzk, deGiorgi:2026jqn, Batell:2026avi}. Extensions of this minimal setup can turn the majoron into a QCD axion, which is the subject of this paper. Related constructions in which the PQ-breaking scale is tied to the seesaw scale, with the same scalar sector generating Majorana masses for the right-handed neutrinos, have also been explored in KSVZ-like models~\cite{Shin:1987xc,Dias:2014osa,Ahn:2015pia,Ballesteros:2016euj,Ballesteros:2016xej,Ballesteros:2019tvf}.

We will begin by constructing the models in their simplest realisation, which, as we will see, can only be made cosmologically viable if the PQ symmetry is broken before the end of inflation and never restored afterwards. Nevertheless, part of the discussion will also carry over to the post-inflationary constructions of \cref{sec:post_inflationary}, which extend the models introduced here.

\subsection{KSVZ-like axions with stable protons}

One way to turn the majoron into an axion is to follow KSVZ and include heavy quark pairs $\left(Q_L^{(i)}, Q_R^{(i)}\right)$ that transform in vector-like, colour-triplet representations under the SM gauge symmetry (hence we call them VLQs, a slight misnomer), but chiral under $U(1)_X$ with charges $(L_i, R_i)$.
They get masses at the scale of $U(1)_X$ breaking due to Yukawa couplings to $\phi_{1,2}$ 
\begin{equation} \label{eq:KSVZ}
    \mathcal{L}_{\rm KSVZ} = \sum_{i=1}^2 \left(\overline{Q}_L^{(i)} \cdot Y^i \cdot Q_R^{(i)}\right) \phi_i\, ,
\end{equation}
which constrains their possible $U(1)_X$ charges. We allow for $N_i$ copies of the $\left(Q_L^{(i)}, Q_R^{(i)}\right)$ pair, and $Y_i$ is an $ N_i \times N_i$ matrix whose indices we have suppressed.

\subsubsection*{Peccei--Quinn symmetry: identifying the physical axion }

With the addition of VLQs, the global symmetry associated to the Goldstone mode can acquire a mixed anomaly with $SU(3)$ colour, which must be non-zero for the Goldstone to solve the strong CP problem by relaxing the QCD theta angle. We here compute this
mixed anomaly coefficient, at first for a generic VLQ spectrum. Afterwards, we will derive explicit anomaly-free spectra of VLQs that do the job.

The first step is to identify the global symmetry current orthogonal to the Goldstone that is eaten by the $U(1)_X$ gauge field.
It is convenient to parametrise via radial ($\rho_i$) and angular ($a_i$) modes of both $\phi_i$:
\begin{equation} \label{eq:phi_expand}
    \phi_i = \frac{v_i+\rho_i(x)}{\sqrt{2}} \exp(i a_i /v_i)\, ,
\end{equation}
We can substitute into the kinetic term for $\phi_i$ to isolate the physical Goldstone:
\begin{equation} \label{eq:SSB}
    \mathcal{L} \supset \sum_{i=1}^2 |(\partial_\mu - ig_X X_i X_\mu)\phi_i|^2  \supset
    \frac{1}{2}\sum_{i=1}^2 (\partial_\mu a_i - g_X X_i v_i X_\mu)^2\, ,
\end{equation}
Perform a rotation of basis
\begin{equation}
    \begin{pmatrix}
        \varphi_X \\ a
    \end{pmatrix}
    = \frac{d_X}{v_X}
    \begin{pmatrix}
        x_1 v_1 & x_2 v_2 \\
        -x_2 v_2 & x_1 v_1
    \end{pmatrix}
    \begin{pmatrix}
        a_1 \\ a_2
    \end{pmatrix},
    \qquad v_X = \sqrt{X_1^2 v_1^2 + X_2^2 v_2^2} \, ,
    \label{eq:physical_2axion}
\end{equation}
where $d_X = \gcd(X_1,\, X_2)$, and $X_i = d_X x_i$.

From~\eqref{eq:SSB}, the mode $\varphi_X$ gets eaten by the $X_\mu$ gauge field, giving it a mass $M_X = g_X v_X$, while the orthogonal combination $a$ is the physical Goldstone, associated with the global PQ symmetry $U(1)_{\text{PQ}}$. 
By canonically normalising the Goldstone $a$, we identify the effective axion decay constant to be 
\begin{equation} \label{eq:va}
  a= \frac{d_X}{v_X}\left(-x_2 v_2 a_1+x_1 v_1 a_2\right) =\frac{v_1 v_2 d_X}{v_X}(-x_2 \theta_1+x_1 \theta_2)\equiv v_a \theta_a  \, ,\quad  v_a \equiv  d_X \dfrac{v_1 v_2}{v_X} \, ,
\end{equation}
with $\theta_i \equiv a_i/v_i$ and $\theta_a=-x_2 \theta_1+x_1 \theta_2$. 
Note that in this definition, $\theta_a$ has a $2\pi$ period on $T^2/U(1)_X \cong S^1$, since $\gcd{(x_1, x_2)}=1$. From here we can compute the mixed Adler-Bell-Jackiw (ABJ) anomaly~\cite{Adler:1969gk,Bell:1969ts} between PQ symmetry and $SU(3)_c$ due to the chiral heavy quarks. The heavy quark Yukawa Lagrangian~\eqref{eq:KSVZ} gives the terms
\begin{equation}
    \mathcal{L}_{\rm KSVZ} \supset \left(\overline{Q}_L^{(1)} \cdot \frac{v_1 Y^1}{\sqrt{2}} \cdot Q_R^{(1)}\right) e^{-i d_X  X_2 \frac{v_2^2}{v_X^2} a/v_a }
    +
    \left(\overline{Q}_L^{(2)} \cdot \frac{v_2Y^2}{\sqrt{2}} \cdot Q_R^{(2)}\right) e^{  id_X  X_1 \frac{v_1^2}{v_X^2} a/v_a } \, .
\end{equation}
The axion can be removed from these terms via a chiral rotation
\begin{equation}
    Q_{L/R}^{(1)} \to e^{\mp i d_X  X_2 \frac{v_2^2}{v_X^2} \frac{a}{2v_a}} Q_{L/R}, \qquad 
    Q_{L/R}^{(2)} \to e^{\pm i d_X  X_1 \frac{v_1^2}{v_X^2} \frac{a}{2v_a}} Q_{L/R}^{(2)} \, .
    \label{eq:field_red_ax}
\end{equation}
This chiral rotation generates the following term coupling the axion to the gluon instanton density,
\begin{equation}
\label{eq:gluon_coupling}
    \mathcal{L}_{\rm axion} = \frac{\mathcal{A}_{\rm PQ}}{32\pi^2} \frac{a}{v_a} g_s^2 G \widetilde{G} \quad \textrm{with} \quad  \mathcal{A}_{\rm PQ} = \frac{d_X}{ v_X^2}\left(N_1 X_2 v_2^2 - N_2 X_1 v_1^2\right) \, ,
\end{equation}
where $g_s$ is the strong coupling,
which, as usual, relaxes the theta term to solve the strong CP problem. Recall that, for $SU(n_c)$ gauge field $G$,  the normalisation of instanton number is such that the integral $\frac{g_s^2}{32\pi^2}\int_M d^4x\, G \tilde{G} \in \Z$ for any compact 4-manifold $M$.

This expression simplifies further for anomaly-free spectra of VLQs. We discuss anomaly cancellation in the next Subsection, but now all we need to know is that the linear anomaly cancellation condition requires $(N_1, N_2) = k(x_2, -x_1)$. This implies
\begin{equation}
\label{eq:pq_ano}
      \mathcal{A}_{\rm PQ} = \frac{N_1}{x_2} = -\frac{N_2}{x_1} = k \in \Z \, ,
\end{equation}
having also substituted in the expression for $v_X$. Reassuringly, the anomaly coefficient is an integer in these units. It is also easy to check that shifting $a \to a + 2\pi v_a$ leaves the partition function phase invariant, verifying that $a/v_a$ indeed has period $2\pi$.

\subsubsection*{Anomaly cancellation}

We now want to characterise which heavy quark spectra are anomaly-free, to complete the EFTs just described.
To begin with (we will generalise this assumption in \S \ref{sec:post_inflationary}), we suppose that the VLQs are colour triplets with zero EW charges, for which the analysis is particularly simple.\footnote{These VLQ representations typically give rise to stable relics. However, since the models introduced in this Section will feature domain walls anyway, we need to presume the $U(1)_X$ condensation occurs before inflation. In that case we do not need to worry about stable relics, because the VLQs will not be produced after inflation provided their masses exceed the reheating temperature. We turn to post-inflationary scenarios in \S \ref{sec:post_inflationary}.} 

The anomaly cancellation conditions (ACCs) here give the Diophantine equations
\begin{align}
    0 = 
    \sum_{i=1}^2 N_i (L_i - R_i), \qquad
    0 = \sum_{i=1}^2 N_i (L_i^3 - R_i^3),
\end{align}
where the assumption that heavy quarks have zero hypercharge means there is no quadratic to solve.
The Lagrangian~\eqref{eq:KSVZ} implies $X_i=L_i-R_i$. It is convenient to change variables to $X_i$ and $S_i=L_i+R_i$, for which the anomaly conditions become
\begin{align}
    0 &= \sum_i N_i X_i, \label{eq:2scalar_linear}\\
    0 &= \sum_i N_i X_i (3 S^2_i + X^2_i) \, , \label{eq:2scalar_cubic}
\end{align}
where we used $a^3-b^3=(a-b)(a^2+ab+b^2)$ to factor out $X_i$ also from the cubic.
The linear condition~\eqref{eq:2scalar_linear} implies $(N_1, N_2) = k (x_2, -x_1)$, as we used in~\eqref{eq:pq_ano} to simplify the PQ anomaly coefficient.

It is quick to see that the cubic reduces to a quadratic on the plane defined by the linear condition.
Assuming also that $X_1, X_2 \neq 0$ and $N_1, N_2 \neq 0$, so that no fields decouple, Eq.~\eqref{eq:2scalar_cubic} becomes
\begin{equation} \label{eq:reduced_quadratic}
    3S_1^2 + X_1^2 = 3S_2^2+X_2^2\,.
\end{equation}
This equation has infinitely many integer solutions, which can be parametrised using a little number theory.\footnote{For completeness, the full space of solutions is parameterised by three integer variables $(\alpha, s_1, s_2)$, as follows:
    $X_1 = \alpha^2 - 3(s_1^2-s_2^2)$,
    $X_2 = \alpha^2 + 3(s_1^2 - s_2^2)$, 
    $S_1 = 2\alpha s_1$, and 
    $S_2 = 2\alpha s_2$.
It can be proven that this is the general solution (see~\cite[\S 3.2]{Allanach:2018vjg}). }
We are not, however, interested in the general solution, but rather in particular solutions for which the scalar charges $X_i$ preserve the discrete, proton-stabilising $\Z_9$ symmetry.
It is convenient to make a further assumption to reduce the quadratic to a linear equation.\footnote{A similar strategy was used to find anomaly-free chiral models in Ref.~\cite{Davighi:2021oel}.} To do so, note that~\eqref{eq:reduced_quadratic} factorises:
\begin{equation}
    (X_1-X_2)(X_1+X_2) = 3(S_2-S_1)(S_1+S_2)\, .
\end{equation}
If then, for instance, we assume 
\begin{equation} \label{eq:PRE-BM-choice}
    S_2 = S_1 + X_2-X_1, \qquad \text{equivalently~~} R_1=R_2\, ,
\end{equation}
then the ACCs reduce to $X_1-2X_2 = 3S_1$. 

The upshot is:
for any $X_1$ and $X_2$ of interest, we can find a solution for $S_i$ and hence for the heavy quark charges $L_i$ and $R_i$, which are integers due to \cref{eq:r2equalsr3}. 

\subsection{Solving the axion quality problem} \label{sec:quality}

The axion quality problem concerns the extent to which we can control other PQ symmetry-breaking effects that would contribute to the pNGB axion potential and could displace its minimum from the CP-restoring point.\footnote{Of course, the interactions~\eqref{eq:explicit_breaking_term} that break PQ symmetry do not run the risk of breaking the discrete $\Z_9$ gauge symmetry that enshrines proton stability, because they are $U(1)_X$ gauge invariant and $\Z_9 \subset U(1)_X$.  } One expects generic global-symmetry-breaking operators to be generated at the very least by quantum gravity effects (given quantum gravity is expected to break all global symmetries), and so we should expect our Lagrangian to feature generic operators allowed by gauge invariance with appropriate Planck-scale suppression. Such is the smallness of the observed value of $|\bar{\theta}|$, that operators up to some high mass dimension ought to be suppressed in order for an axion to provide a quality solution to the strong CP problem.

\subsection*{Tree-level scalar potential}

In our setup, the $U(1)_X$ gauge symmetry controls which operators can be written down in the scalar potential using the fields $\phi_1$ and $\phi_2$. 
We seek charge assignments that forbid PQ-violating operators up to some mass dimension $d_{\text{PQ}}$ in order to preserve axion quality. 
The leading Planck-suppressed operator appearing at tree-level in the scalar potential can be written as~\cite{Greljo:2025suh} 
\begin{align} \label{eq:explicit_breaking_term}
     V_\textrm{grav}&= \frac{\eta}{M_\textrm{Pl}^{|s|+|t|-4}}\phi_1^{[s]} \phi_2^{[t]} + \textrm{H.c.} \,,\quad
    \text{where} \quad \phi_i^{[s]} = \begin{cases}
        \phi_i^{|s|} & \text{for } s\geq 0, \\
        (\phi_i^\ast)^{|s|} & \text{for } s< 0,
    \end{cases}
\end{align}
where $M_\textrm{Pl}=1.22\times 10^{19}\,$GeV is the Planck mass\footnote{Note that to compare with the usual conventions, we have taken a less conservative power-counting than NDA would estimate~\cite{Martucci:2024trp}.}, and $\eta$ is a complex coefficient that could be $\mathcal{O}(1)$. Using gauge invariance,
\begin{equation}
    s X_1 + t X_2 = 0
    \qquad\Longrightarrow\qquad
    (s,t)=\ell(x_2,-x_1)\,,
    \qquad \ell\in\mathbb Z\, ,
\end{equation}
where we have used $X_i=d_X x_i$ and $\gcd(x_1,x_2)=1$. The leading operator will always be for $\ell=1$. Hence 
\begin{equation}
    d_{\text{PQ}} = |x_1|+|x_2|\, ,
\end{equation}
is determined by the scalar charges under $U(1)_X$, and
the leading gravitational contribution to the axion potential then takes the form 
\begin{align}
\label{eq:gravity_potential}
    V_{\rm grav}
    &=
    2|\eta| M_{\rm Pl}^4
    \left(\frac{v_1}{\sqrt{2}M_{\rm Pl}}\right)^{|s|}
    \left(\frac{v_2}{\sqrt{2}M_{\rm Pl}}\right)^{|t|}
    \cos\left(
        \alpha_\eta-\frac{a}{v_a}
    \right)\, .
\end{align}
where $\alpha_\eta \equiv \arg\eta$. To get an idea for the strength of this constraint, let us take the scales $v_1,\, v_2$ associated with the breaking of $U(1)_X$ symmetry (and thus also PQ) to be at least $10^9$ GeV. For a generic order-1 phase, this then requires\footnote{When we construct explicit models, we will calculate the axion quality bound on a case-by-case basis, as a function of varying parameters $v_1$ and $v_2$. Typically, viable models will require $d_{\rm PQ}$ larger than 10. } 
\begin{equation} \label{eq:quality}
    d_{\text{PQ}} = |x_1|+|x_2| \geq 10\, ,
\end{equation}
to not shift $\bar{\theta}$ by more than $
\sim 10^{-10}$. 

\subsection*{Radiative contributions from the heavy quarks}

It is not only pure scalar operators that can violate the PQ symmetry. Our models all feature a spectrum of heavy VLQs charged chirally under $U(1)_X$. The $U(1)_X$ gauge symmetry might permit operators involving fermions and scalars that violate PQ global symmetry; if the fermions can be closed in loops, this can give radiative contributions to the scalar potential that might also endanger axion quality~\cite{Bonnefoy:2022vop}. To grant a full solution to the axion quality problem, we should also ensure these effects are sufficiently small. 

Consider a gauge-invariant effective operator of the form
\begin{equation} \label{eq:PQV-op-fermions}
    \mathcal{L} \supset \frac{c}{M_{\text{Pl}}^{3k+|s|+|t|-4}}\prod_{r=1}^k \left( \overline{Q}_L^{(i_r)}   Q_R^{(j_r)}\right) \phi_1^{[s]} \phi_2^{[t]} \, ,
\end{equation}
with net PQ charge.\footnote{The PQ charge of the VLQs can be deduced from their couplings to the scalars $\phi_i$.} This form covers chirality-changing bilinears, but operators may also contain same-chirality currents such as $(\overline Q_{L,i}\gamma_\mu Q_{L,j})(\overline Q_{L,k}\gamma^\mu Q_{L,l})$ which must be checked separately.

If there exists a valid graph allowing one to close the fermion line via renormalisable vertices, this can give a $k$-loop contribution to the PQ-violating potential, of expected size
\begin{equation}
\label{eq:grav_ferm_loop}
    \Delta V_{\text{grav}}^{(k)} \sim  c\mathcal{Z} \left(\frac{n_c}{16 \pi^2}\right)^k 
    \frac{M^{3k}_{Q} \phi_1^{[s]} \phi_2^{[t]}  }{M_{\text{Pl}}^{3k+|s|+|t|-4}}\, ,
\end{equation}
where $n_c$ is the color factor and the dimensionless $\mathcal{Z}$ denotes the relevant Yukawa, mixing, flavor, and contraction factors. Conservatively, we take $c \mathcal{Z} \sim 1$.
If such operators exist at a lower mass dimension than the leading PQ-violating scalar operator discussed above, they can give the leading PQ-violation (depending on the various scale ratios and dimensions). We can impose a conservative condition to preclude such effects, which is to require {\em all} closure-relevant operators of form similar to~\eqref{eq:PQV-op-fermions} have mass dimension $\geq d_{\text{PQ}}$, the leading operator dimension appearing in the scalar potential. Then, the ratio of a contribution with $k$ fermion bilinears relative to the pure-scalar operator is
\begin{equation}
    \frac{\Delta V_{\text{grav}}^{(k)}}{\Delta V_{\text{grav}}^{(0)}} \sim \left(\frac{n_c}{16\pi^2} \right)^k \left(\frac{M_{Q}}{v_i}\right)^{3k}\, ,
\end{equation}
which is subleading for VLQ masses at the PQ breaking scale, $M_Q \sim v_i$. This condition will be very important later in \S \ref{sec:filterPQ} when we come to construct post-inflationary models.

The VLQ spectrum might be such that there is in fact no valid Feynman graph allowing us to close the loop and shift the potential. In that case, the heavy-fermion sector cannot provide a lower-dimensional shortcut to PQ breaking, and we do not spoil axion quality.

For our pre-inflationary solutions~\eqref{eq:PRE-BM-choice}, the chosen anomaly-free branch satisfies
\begin{equation}
 R_1=R_2\,,\qquad L_i-R_j=X_i\quad(i,j=1,2)\, .
\end{equation}
Consequently, every entry of the heavy-quark mass matrix is generated by one scalar,
\begin{equation}
 \mathcal M\sim
 \begin{pmatrix}
 Y_{11}\phi_1&Y_{12}\phi_1\\
 Y_{21}\phi_2&Y_{22}\phi_2
 \end{pmatrix},
\end{equation}
and no dimension-three mass term is allowed. Closing each of $k$ chirality-changing bilinears $\overline Q_{L,i}Q_{R,j}$ therefore requires one scalar mass insertion. Replacing them by the corresponding signed scalar spurions maps an operator with $n_\phi$ explicit scalars onto a PQ-violating pure-scalar operator with $n_\phi+k$ scalars. Hence
\begin{equation}
 n_\phi+k\geq d_{\rm PQ}
 \qquad\Longrightarrow\qquad
 D_{LR}=3k+n_\phi\geq d_{\rm PQ}+2k\, ,
\end{equation}
where $D_{LR,JJ}$ denotes the dimension of a gauge-invariant operator that contains $LR$ or $JJ=LL,RR$ fermion currents in addition to scalars.
Thus chirality-changing operators cannot dominate the leading scalar contribution.

Since $R_1=R_2$, all right-handed currents are gauge neutral and may be chosen PQ neutral. It remains only to consider left-handed currents. Write the primitive scalar charges as $(x_1,x_2)=(-b,a)$ with positive $a,b$, so that $d_{\rm PQ}=a+b$. If $c$ denotes the multiplicity of the off-diagonal current $\overline Q_{L,1}\gamma_\mu Q_{L,2}$ in a composite operator $\mathcal J_c$, gauge invariance of $\phi_1^{[p]}\phi_2^{[q]}\mathcal J_c$, requires
\begin{equation}
 -bp+aq+c(a+b)=0
 \qquad\Longrightarrow\qquad
 p=c+a\ell\,,\quad q=b\ell-c\,,\quad \ell\in\mathbb Z\, .
\end{equation}
The operator violates PQ only for $\ell\neq0$, and hence
\begin{equation}
 |p|+|q|\geq|p+q|=|\ell|(a+b)\geq d_{\rm PQ}\, .
\end{equation}
A derivative-free Lorentz scalar contains at least two vector currents, so $D_{JJ}\geq d_{\rm PQ}+6$.

As an example, for the chosen benchmarks in \cref{tab:benchmark_models}, we find
\begin{equation}
 d_{\rm PQ}=(10,13,17),\qquad
 D^{LR}_{\min}=(12,15,19),\qquad
 D^{JJ}_{\min}=(16,19,23)
\end{equation}
for benchmarks (Pre-1, Pre-2, Pre-3) respectively. Thus, the potentially dangerous radiative shortcut is absent in all three pre-inflationary benchmarks. Operators with non-zero heavy-quark number cannot close at a single insertion and do not alter this conclusion.

Mixed operators involving SM fermions or RH neutrinos do not change this conclusion. SM Yukawa closures insert the PQ-neutral Higgs, while RH-neutrino mass closures use the same scalar spurions $\phi_{1,2}$, so they reduce to the scalar analysis above. In particular, the allowed terms $ N_S N_D \phi_1$ and $N_D N_D\phi_2$ fix $q_{N_S}+q_{N_D}+q_{\phi_1}=0$ and $2q_{N_D}+q_{\phi_2}=0$. Hence the gauge-invariant dimension-six operator $\overline N_S^c N_S\phi_1^2\phi_2^\ast$ is automatically PQ preserving.

Together, these ingredients provide a logical path to building models in which $U(1)_X$ ensures exact proton stability as well as sufficient axion quality. 
Choose coprime $(x_1, x_2)$ such that $-x_1 + x_2 = d_{\text{PQ}}$ (we consider solutions with positive $x_2$ and negative $x_1$.). For any such pair, we can construct an anomaly-free spectrum of VLQs; the primitive signed multiplicities $(N_1,N_2)=(x_2,-x_1)$ construct an anomaly-free spectrum containing precisely $n_{\text{VLQ}}=N_1+N_2=d_{\text{PQ}}$ heavy-quark representations. 

We next discuss several other cosmological and laboratory constraints on our framework, before defining a set of specific benchmark models with which we explore the parameter space.

\subsection{Cosmological abundance and dark matter}
\label{sec:DM}

Apart from solving the strong CP problem, the QCD axion also provides a natural dark-matter candidate. In the early Universe, while the Hubble expansion rate is larger than the axion mass, the axion field is overdamped and remains approximately frozen at some initial value $a_i=f_a\theta_i$. From now on, we identify $f_a \equiv v_a$, with $v_a$ defined in \cref{eq:va}. Once the temperature decreases sufficiently that $H\sim m_a(T)$, the field starts oscillating around the minimum of its potential. These coherent oscillations behave as non-relativistic matter, with an energy density that redshifts as $\rho_a\propto a^{-3}$, and can therefore constitute cold dark matter through the misalignment mechanism~\cite{Preskill:1982cy,Abbott:1982af,Dine:1982ah,DiLuzio:2020wdo}.

The resulting abundance depends on the initial misalignment angle $\theta_i$. It can be written schematically as
\begin{equation}
    \Omega_a h^2 \simeq
    0.12\,
    \left(\frac{\theta_i}{2.15}\right)^2
    \left(
        \frac{f_a}{2\times10^{11}\,\textrm{GeV}}
    \right)^{7/6}\,,
\end{equation}
where $\theta_i$ depends on the relative time of the PQ breaking and the end of inflation. In the pre-inflationary scenario, the PQ symmetry is broken during inflation and is not restored afterwards. Inflation then selects a single initial misalignment angle $\theta_i$ across our observable Universe, which remains a free parameter controlling the axion relic abundance. In the post-inflationary scenario, the axion abundance becomes approximately predictive in terms of the axion scale $f_a$, because different causally disconnected regions of the Universe sample random initial misalignment angles. Averaging over these regions fixes the typical contribution from the effective angle $\langle\theta_i^2\rangle\simeq 2.15^2$ after including anharmonicity~\cite{GrillidiCortona:2015jxo}, leaving $f_a$ as the main parameter controlling the relic abundance~\cite{DiLuzio:2020wdo}. This opens the well-known possibility that the QCD axion could constitute the DM, for $f_a \approx 2 \times 10^{11}$ GeV. On the other hand, if $f_a$ exceeds this, the energy density in the axion field exceeds that in DM, contrary to observations. This constrains the models, which we include in Fig.~\ref{fig:post_inf}. The pre-inflationary scenario provides more freedom to fit the relic abundance because the initial misalignment angle is a priori undetermined; in particular, larger $f_a$ is enabled for smaller $\theta_i$. 

An additional contribution to the axion DM abundance in post-inflationary scenarios arises from axion radiation by the cosmic-string network. This contribution is parametrically comparable to misalignment, although its precise size must be determined from numerical simulations. Recent simulations typically find that string radiation shifts the preferred QCD-axion mass into the range $ m_a \simeq 40-450\,\mu{\rm eV}$ ~\cite{Saikawa:2024bta, Benabou:2024msj}, while remaining within the same order of magnitude as the misalignment-only prediction. Consequently, including string radiation would generally shift the dark-matter-favoured region towards smaller values of $f_a$, and therefore towards smaller values of both symmetry-breaking VEVs. This can potentially open, or enlarge, the viable parameter space. Since the string network in our models differs from the standard axion case, the existing simulation results cannot be applied directly. We nevertheless use them as a rough estimate of the possible size of the string-radiation contribution, while keeping in mind that the subsequent string evolution is different in our setup. 

At distances larger than the radial and gauge-field cores, the strings behave as ordinary global axion strings, since the gauged combination is screened and only the physical axion remains massless. For a string with axion winding $w_a$, its infrared tension is therefore
\begin{equation}
    \mu_a^{\rm IR}\simeq \pi f_a^2 w_a^2 \ln\left(\frac{L}{\delta}\right),
\end{equation}
with $L\sim H^{-1}$ and $\delta$ the core size. If the network reaches scaling, its axion-radiation rate is thus expected to have the same parametric form as in the standard post-inflationary scenario, up to order-one differences in the scaling density, winding distribution, and core contribution. In particular, when the string decomposition occurs close to the QCD epoch, $t_2\sim t_{\rm QCD}$, we expect the resulting string contribution to the axion abundance to be of the same order as in the conventional $N_{\rm DW}=1$ case. This expectation can fail for sufficiently delayed decomposition or large hierarchies, where the subsequent string-wall dynamics may significantly modify the axion production and a dedicated simulation is required~\cite{Mupo:2025ner}.

\subsection{Inflation, isocurvature bounds and leptogenesis}

As seen above, the cosmological history, and in particular the time of inflation relative to PQ breaking, is crucial in determining the axion DM abundance. Moreover, it also determines the cosmological bounds that must be imposed on our models. If the PQ symmetry is broken after inflation, or is restored by the thermal bath after reheating, the heavy quarks can be thermally populated. If these states are sufficiently long lived, their abundance may survive until late times and their decays can inject energetic SM particles into the plasma, potentially disrupting the successful predictions of BBN. Furthermore, PQ breaking after inflation leads to the formation of topological defects. Their subsequent evolution and decay must occur sufficiently early to avoid an unacceptable relic abundance or a modification of BBN. We discuss these constraints in \cref{sec:post_inflationary}.

If the PQ symmetry is broken before or during inflation and is never restored afterwards, our observable Universe originates from a single causally connected patch of the PQ vacuum manifold. Any strings produced when the symmetry was broken are exponentially diluted outside the observable horizon, and no new strings are generated after inflation since there is no subsequent PQ phase transition. The axion field therefore has an approximately homogeneous initial misalignment angle across our observable patch, up to inflationary fluctuations. When the QCD potential turns on, the field relaxes everywhere towards the same minimum, so no domain walls are formed. In this case, the heavy relics need not be thermally populated, and the constraints associated with their late decays and with the string-domain-wall network can be avoided. 

This pre-inflationary scenario is nevertheless subject to the usual axion isocurvature constraint and to the requirement that reheating does not restore the PQ symmetry, which we discuss next. Furthermore, in our models, we may also require that the lightest RH neutrino is in thermal equilibrium to account for the BAU via thermal leptogenesis. For a related discussion in majoron models, see Ref.~\cite{Batell:2026avi}.

First, quantum fluctuations of the axion during inflation generate an isocurvature component constrained by the CMB. Second, if thermal leptogenesis is to take place after inflation, the reheating temperature must be sufficiently high to populate the lightest RH neutrino, while remaining below the temperature at which the PQ symmetry would be thermally restored. For an axion that is light during inflation, its fluctuations are of order
\begin{equation}
\delta a \simeq \frac{H_I}{2\pi}\, ,
\end{equation}
where $H_I$ denotes the Hubble scale during inflation. If the axion constitutes a fraction $\xi_a\equiv\Omega_a/\Omega_{\rm DM}$ of the dark matter, the corresponding isocurvature power spectrum is, in the small-misalignment-angle approximation,
\begin{equation}
\mathcal P_S\simeq\xi_a^2\left(\frac{H_I}{\pi f_a^I\theta_i}\right)^2 ,
\end{equation}
where $f_a^I$ denotes the axion decay constant during inflation. The CMB bound measured by the Planck collaboration~\cite {Planck:2018jri} on uncorrelated isocurvature therefore translates approximately into
\begin{equation}
H_I\lesssim2.9\times10^{-5}\,\frac{\theta_i f_a}{\xi_a}\, ,
\label{eq:iso_bounds}
\end{equation}
assuming the VEVs do not change appreciably between inflation and today.
The inflationary scale also places an upper limit on the reheating temperature. From energy conservation, a thermal bath is reheated to temperatures of
\begin{equation}
T_{\rm RH}\lesssim\left(\frac{90}{\pi^2g_\ast}\right)^{1/4}\sqrt{ M_{\rm Pl}H_{\rm end}}\simeq 0.55\sqrt{ M_{\rm Pl}H_I}\, ,
\label{eq:trh_up_bounds}
\end{equation}
where in the last expression we have taken $g_\ast\simeq100$ and $H_{\rm end}\lesssim H_I$. This should be regarded as the maximally efficient, approximately instantaneous-reheating limit; a less efficient reheating phase gives a lower $T_{\rm RH}$ for the same inflationary scale. 

However, leptogenesis requires the lightest RH neutrino to be thermally populated, $T_{\rm RH}\gtrsim M_{N_1}\, $.
The existence of a consistent pre-inflationary thermal history therefore requires a reheating window
\begin{equation}
M_{N_1}
\lesssim T_{\rm RH}
<
\min\left[
T_{\rm PQ}^{\rm rest},
\,
0.55\sqrt{ M_{\rm Pl}
H_I^{\rm iso}}
\right] ,
\label{eq:pre_window}
\end{equation}
where if we assume $v_2>v_1$, the temperature at which the PQ-symmetry is thermally restored can be given by $T_{\rm PQ}^{\rm rest} =\mathcal{O}(v_1)$.

\subsection{Axion bounds and searches}
\label{sec:axion_searches}

Axion couplings can be probed in astrophysical sources, since axion emission provides an additional channel for energy loss. In particular, observations of neutron-star cooling constrain the axion-nucleon coupling, while dedicated helioscope experiments probe the axion-photon coupling through axions produced in the Sun.

If axions constitute the dark matter abundance of the Universe, they can also be targeted by haloscope experiments, which search for the conversion of ambient Galactic axion dark matter into photons in the presence of a strong magnetic field.

For our model, the most important dark-matter-independent bound is the one obtained from neutron-star cooling~\cite{Buschmann:2021juv}, leading to 
\begin{equation}
\label{eq:neutron_star}
  f_a\gtrsim3.6\times 10^8\,\textrm{GeV}\, , \quad  m_a \lesssim 16 \, \textrm{meV} \, .
\end{equation}
Future helioscopes such as IAXO~\cite{IAXO:2019mpb}, will be competitive with neutron star bounds setting bound on the photon coupling of $g_{a\gamma}\lesssim 4.3 \times 10^{-12}\,\textrm{GeV}^{-1}$, which means a bound on the scale of $f_a\gtrsim 5.2\times 10^8\,\textrm{GeV}$, for models with $E/N=0$. The helioscope measurements, which are ultimately expected to find comparable sensitivity, will be subject to different modelling uncertainties to the neutron star cooling bounds, giving an important cross-check.

More promising are haloscope searches, where, if the axion constitutes the full dark-matter relic abundance, a significant part of the QCD axion parameter space can be tested in the upcoming years. In particular, ADMX has already started probing the region predicted by the QCD axion~\cite{ADMX:2025vom}. Together with MADMAX~\cite{Caldwell:2016dcw}, the two experiments are expected to cover approximately the mass range $m_a \simeq 0.7-200\,\mu{\rm eV}$~\cite{Stern:2016bbw,Beurthey:2020yuq}. This region lies in the predicted parameter space where post-inflationary axions could constitute the total amount of DM. 

Finally, while coupling to neutrinos would distinguish this setup from standard KSVZ axion models, it is unfortunately not measurable in the foreseeable future, as the testable seesaw part would require larger masses, see Fig.~7 of Ref.~\cite{Greljo:2025suh}. However, residual axion couplings and low-energy parameters of the EFT could help in resolving this ambiguity~\cite{Palavric:2026vej,Biggio:2026gcs}.

\subsection{Parameter space of pre-inflationary benchmark models}
\label{sec:preinf_pheno}

In this section, we explore the phenomenology of the pre-inflationary scenario. In this case, the VLQs can be chosen to be neutral under hypercharge, avoiding the need to impose any additional anomaly-cancellation conditions beyond the linear and cubic equations in \cref{eq:2scalar_linear,eq:2scalar_cubic}. We therefore take throughout this section $Q\sim (\bm{3},\bm{1})_0\, $. Since these quarks are neutral under the electroweak gauge group, they do not contribute directly to the electromagnetic anomaly. The axion-photon coupling is therefore generated only by the model-independent axion-pion mixing contribution~\cite{GrillidiCortona:2015jxo},
\begin{equation}
    \mathcal{L}_{a\gamma}=\frac{g_{a\gamma}}{4}F\widetilde{F}\quad \textrm{with}\quad  g_{a\gamma}= -\frac{\alphaem}{2\pi}\frac{1.92(4)}{f_a} \, .
\end{equation}

To explore the large phenomenological landscape of models, we consider a set of representative Benchmark Models (BMs). The neutrino texture introduced in \cref{sec:neutrino_texture} already leads to a non-trivial range of possible PQ qualities. The lowest-dimensional gauge-invariant operator that explicitly breaks the
accidental PQ symmetry therefore has dimension
\begin{equation}
    d_{\rm PQ}=|s|+|t| =|x_1|+|x_2|=\frac{|r_{S}+r_{D}|+2|r_{D}|}{g}\, .
    \label{eq:dpq_texture}
\end{equation}
For primitive lepton charges, $\gcd(r_{S},r_{D})=1$, one has $g=1$ or $g=2$. The first case corresponds to $U(1)_X\to\mathbb Z_9$ and is realised when $r_{S}$ and $r_{D}$ have opposite parity. In this case, $d_\PQ $ is necessarily odd. Instead, if both $r_{S}$ and $r_{D}$ are odd, $g=2$ and the scalar VEVs leave an unbroken $\mathbb Z_{18}$ subgroup. In this latter case, the additional common factor of two in the scalar charges is removed when defining the primitive PQ charges. This second branch therefore also allows even values of $d_{\rm PQ}$ while retaining the proton-stabilising $\mathbb Z_9$ as a subgroup.

\begin{table}[t]
\centering
\begin{tabular}{c c c c c c c c}
\toprule
BM & $d_{\rm PQ}$ & $(r_{S},r_{D})$ & $m$
& $(x_1,x_2)$ & $(X_{\phi_1},X_{\phi_2})$
& Residual sym. & PQ-brk. \\
\midrule

Pre-1 & 10 & $(-13,7)$   & $1$   & $(-3,7)$   & $(-54,126)$
& $\mathbb Z_{18}$
& $\phi_1^7\phi_2^3$ \\

Pre-2 & 13 & $(-10,3)$     & $-4$  & $(-7,6)$  & $(-63,54)$
& $\mathbb Z_9$
& $\phi_1^6\phi_2^7$ \\

Pre-3 & 17 & $(-12,5)$ & $-2$ & $(-7,10)$ & $(-63, 90)$
& $\mathbb Z_{9}$
& $\phi_1^{10}\phi_2^7$ \\
\bottomrule
\end{tabular}
\caption{Representative benchmark models. The charges $(x_1,x_2)$ denote the primitive scalar charges, while $(X_{\phi_1},X_{\phi_2})$ are the corresponding $U(1)_X$ gauge charges. The last column gives the leading gauge-invariant operator that explicitly breaks the accidental PQ symmetry, up to Hermitian conjugation.}
\label{tab:benchmark_models}
\end{table}

\begin{figure}[tb]
    \centering
    \includegraphics[width=\linewidth]{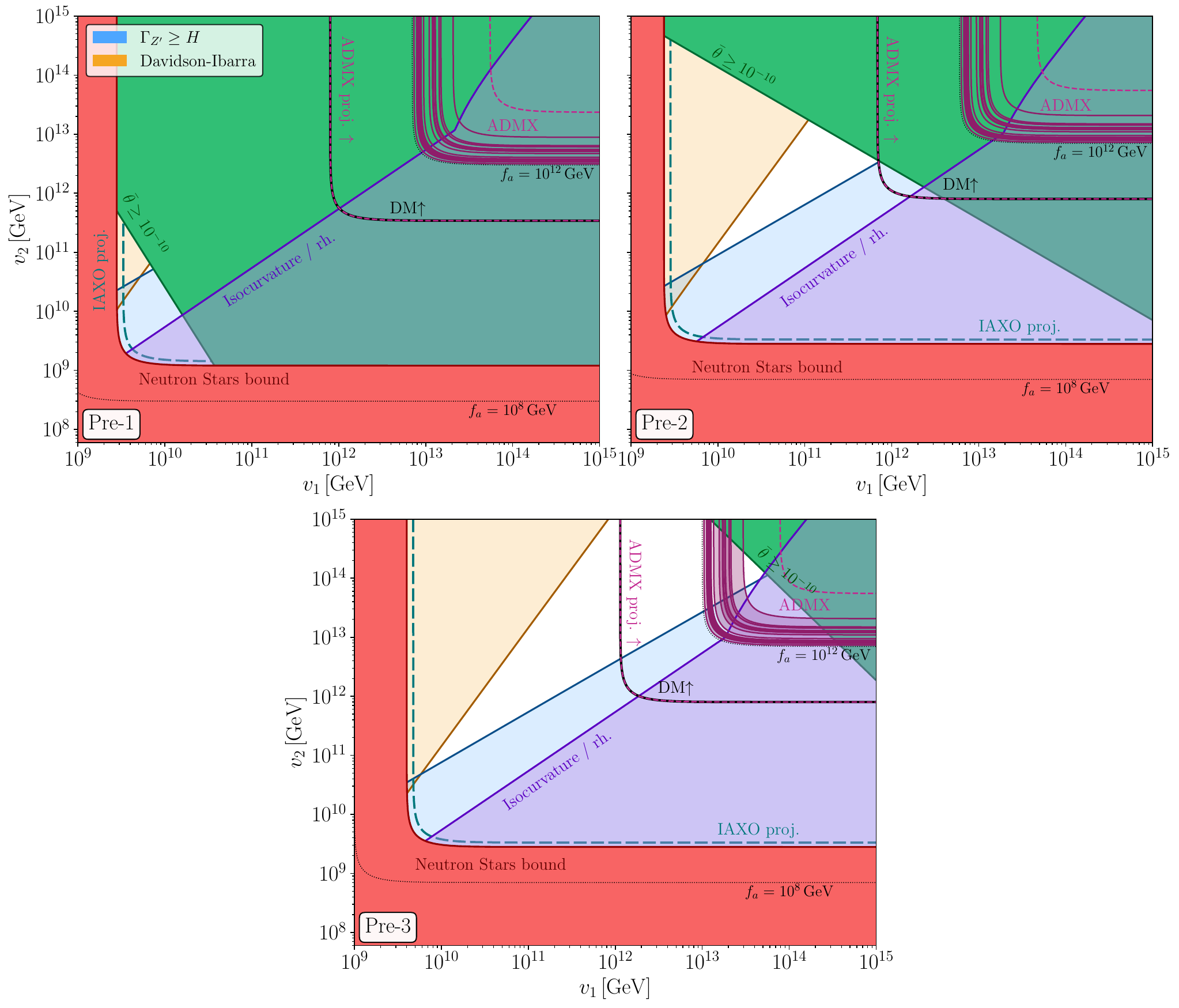}
    \caption{
    Phenomenology of the benchmark models in the pre-inflationary scenario, shown in the $(v_1,\,v_2)$ plane. The PQ-quality bound from $\bar\theta$ is shown in green. Optional leptogenesis constraints are displayed with lower opacity: the DI bound in orange and the decoupling of gauge interactions in blue. Isocurvature and reheating constraints are shown in purple. Dashed contours indicate the projected reach of IAXO and ADMX, while the red region is excluded by neutron-star cooling. The black contour marks the minimum $f_a$ required for axions to account for all of the dark matter, and $\theta_i$ is fixed to reproduce the total DM abundance.
    }
    \label{fig:pre_inf_models}
\end{figure}

In \cref{tab:benchmark_models} we select representative models spanning both branches and several values of $d_{\rm PQ}$. These benchmarks illustrate the interplay of the different phenomenological constraints. The constraints for the three benchmark models in the pre-inflationary scenario are shown in \cref{fig:pre_inf_models}. We see that lower values of the $v_i$ are bounded by neutron-star cooling constraints, \cref{eq:neutron_star}. For each model, we also observe the characteristic behaviour of the PQ-breaking bound, whose slope depends on the relative powers of $\phi_1^{[s]}\phi_2^{[t]}$, see \cref{tab:benchmark_models}. We find that lowest PQ-quality allowed model compatible with the bounds is Pre-1 with $d_\PQ=10$. Interestingly, this number of VLQs preserves the asymptotic freedom of QCD, while just allowing for a compatible solution to the axion quality problem. 
We discuss the running of the SM gauge couplings more in \cref{sec:Landau_Poles}.

We also see from \cref{fig:pre_inf_models} that Pre-1 cannot account for the full DM abundance through misalignment, and that explaining the relic abundance requires $d_\PQ\geq 13$. Even more noteworthy, despite having only $d_\PQ=N_Q=10$ and therefore preserving QCD asymptotic freedom, Pre-1 can satisfy the axion-quality requirement while retaining a small window compatible with the leptogenesis constraints.

Models Pre-2 and Pre-3 can account for the DM abundance, with Pre-2 having a very small window compatible with leptogenesis constraints. Constraints from isocurvature and thermal restoration of the symmetry bound the lower right corner of all models, being less important than the constraint of $\Gamma_X/H\gtrsim 1$. We also see that current constraints from ADMX~\cite{ADMX:2018gho,ADMX:2019uok,ADMX:2021nhd,ADMX:2024xbv,ADMX:2025vom}, lay above the quality constraint for all these BMs, however the projected sensitivity~\cite{Stern:2016bbw} could sweep the DM area compatible with leptogenesis for these models.

Although one could extend the region allowed by quality, and hence the viable DM region, going to larger values of $f_a$ requires increasingly small values of $\theta_i$, which would in turn require further justification, for instance through sufficiently low-scale inflation~\cite{Graham:2018jyp,Takahashi:2018tdu}. Hence, although quality can be further increased, going to models with greater number of quarks $d_\PQ=N_Q$, it is not phenomenologically necessary with current constraints.

\section{Post-inflationary models}
\label{sec:post_inflationary}

As shown in Ref.~\cite{Greljo:2025suh}, the lepton sector of these models can explain the origin of the baryon asymmetry of the Universe via leptogenesis. For this to happen, it is necessary that the RH neutrinos reach thermal equilibrium with the SM. This can be achieved in pre-inflationary models, as we saw in the previous section, by choosing the reheating temperature such that the potential is not thermally restored while the lightest RH neutrinos are, which perhaps would require a coincidence. Another possibility of making the model cosmologically viable in the scenario where the PQ symmetry is broken after inflation. In this scenario, however, major cosmological issues must be addressed: topological defects and stable relics.

\subsection{The domain wall problem for axions}
\label{sec:DomainWalls}

If the PQ symmetry-breaking transition occurs after inflation, there is a well-known domain wall problem, whereby a large amount of energy density is stored in a domain wall network that does not dilute away. The basic reason for this is as follows.

In addition to the spontaneous breaking of PQ symmetry by the scalar condensate, the mixed anomaly between $U(1)_{\text{PQ}}$ and $SU(3)_c$ also provides an explicit breaking of the global PQ symmetry:
\begin{equation}
    \text{ABJ anomaly:} \quad U(1)_{\text{PQ}} \to \Z_{N_\text{DW}}\,.
\end{equation}
This discrete subgroup is given precisely by the anomaly coefficient. For our models,
\begin{equation}
    N_\text{DW} = \mathcal{A}_{\text{PQ}} = k \, ,
\end{equation}
using~\eqref{eq:pq_ano}.
A non-trivial discrete group $\Z_{N_{\text{DW}}}$ would imply there are $N_{\rm{DW}}$ 
degenerate minima of the axion potential, and causally-disconnected patches of the Universe could settle into different vacua separated by domain walls. 
A cosmological problem is avoided if~\cite{Sikivie:1982qv,Vilenkin:1982ks} 
\begin{equation} \label{eq:noDW_naive}
\NDW=|\A_\PQ|=1\, .
\end{equation}
Satisfying the condition~\eqref{eq:noDW_naive} does not mean that domain walls never form in the early Universe. Global strings are first produced at the PQ phase transition through the Kibble mechanism~\cite{Kibble:1976sj}. At later times, when the QCD-induced axion potential becomes relevant, domain walls form and attach to these strings. What $N_{\rm DW}=1$ guarantees is that each string is attached to a single domain wall. In this case the wall tension pulls on the strings, rendering the string-wall network unstable and causing it to collapse~\cite{Sikivie:1982qv,Vilenkin:1982ks}.

\subsection*{Review: domain walls in two-scalar models}

However, for the models presented in this work, which are like those first considered by Barr and Seckel~\cite{Barr:1992qq}, the strings formed in the early Universe have a more complex structure. The physical axion here arises as a linear combination of the phases of two complex scalar fields, which means that a microscopic string is characterised by two integer winding numbers,
\begin{equation}
\Delta\theta_i=2\pi n_i\,,
\qquad (n_1,n_2)\in \Z \times \Z \, .
\end{equation}
The winding of the physical axion around such a string is instead given by
\begin{equation} \label{eq:winding}
    w_a(n_1,n_2) = x_2 n_1 - x_1 n_2\,.
\end{equation}
For the primitive anomaly solution with anomaly coefficient $A_\PQ=1$, the number of QCD domain walls that attach to this particular string is
$\NDW(n_1,n_2)=\left|x_2n_1-x_1n_2\right|$.
Thus, the fact that the theory has a single inequivalent QCD vacuum does not imply that every microscopic string carries a single unit of axion winding.

The consequences of this complication for whether or not stable string-wall networks will form in these models has been discussed extensively in the literature. Here we review some important arguments, before justifying the condition we shall adopt in our study (and noting its caveats).

Barr and Seckel proposed that a stable string-wall network can be avoided on energetic grounds~\cite{Barr:1992qq}, if there is a hierarchy between symmetry breaking scales, $v_2 > v_1$. (Recall that, for our setup, a mild hierarchy in scales is indeed required by leptogenesis, as per~\eqref{eq:leptogen}.) The corresponding phase transition produces the fundamental $\phi_2$ strings, with $n_2=1$. When $\phi_1$ subsequently condenses, its phase can acquire an integer winding $n_1$ around the pre-existing string. The energetically preferred configuration is therefore the one that minimises $w_a|_{n_2=1}$.

If there exists an integer $n_1$ satisfying the final composite string carries a single unit of physical axion winding. Once the QCD potential turns on, only one domain wall attaches to this string and the resulting string-wall network is expected to collapse, as in the standard $N_{\rm DW}=1$ axion scenario. In our models it is possible to realise the one-wall configurations of Barr and Seckel. For example, for the models~\eqref{eq:r2r3} with partial universality this requires $r_{S}+r_{D}=\pm1\pmod{2r_{D}}$ be satisfied, selecting certain patterns for the leptonic couplings of the model.

However, it was recently pointed out in Ref.~\cite{Lu:2023ayc} that this kinematic solution lacks a dynamical mechanism that would guarantee the formation of the energetically preferred strings. In particular, the Kibble mechanism is local: causally disconnected regions choose their scalar phases independently, and there is no reason for the winding of the second scalar around every pre-existing string to arrange itself so as to minimise the asymptotic gradient energy. The resulting network may therefore contain strings with winding numbers different from the Barr--Seckel minimum, with more than one QCD domain wall attached to them. Thus, the existence of an energetically favoured one-wall string configuration does not by itself guarantee that the cosmological string network is composed only of such strings.

Mupo and Zhang since proposed a possible dynamical resolution to this problem~\cite{Mupo:2025ner}. They showed that a general string with non-unit axion winding $|w_a(m,n)|>1$ can nevertheless be decomposed into a number of unit-winding axion strings plus pure gauge strings, {\em i.e.} we can always write\footnote{A string with unit winding $w_a=1$ has winding numbers satisfying $x_2 n_1^\ast - x_1 n_2^\ast=1$ (the existence of such a pair $n_i^\ast$ is guaranteed for our models by B\'ezout's identity since $\gcd(x_1,x_2)=1$; equivalently, this is the condition $N_{\text{DW}}=k=1$).
Pure gauge strings have winding numbers $(n_1^g,n_2^g) \propto (x_1, x_2)$, so that $w_a=0$ using~\eqref{eq:winding}. Now, a unit-winding string and a pure gauge string form a basis for the integer winding lattice, {\em i.e.} for all possible strings. Hence, any string can be decomposed into a number of unit strings pure a number of pure gauge strings~\cite{Mupo:2025ner}. } 
\begin{equation}
    (m,n)=w_a(m,n)\,\underbrace{(n_1^\ast,n_2^\ast)}_{\text{unit string}} + c\,\underbrace{(x_1,x_2)}_{\text{pure gauge}}\,.
\label{eq:string_decomposition}
\end{equation}
Therefore, even if higher-winding strings are produced in the early Universe, one may ask whether they are likely (energetically speaking) to decay into unit-winding strings before the domain-wall network becomes cosmologically dangerous. But it turns out the spontaneous decomposition can be energetically obstructed~\cite{Mupo:2025ner}: the tension of a daughter configuration (unit strings plus pure gauge strings) can exceed that of the parent string, the latter being $\mu_{(m,n)}  \simeq 2\pi n^2 v_2^2$ in the limit $v_2\gg v_1$ relevant for the (leptogenesis-compatible) models we consider. This prevents spontaneous decomposition, in the absence of some additional energy injection.

An additional source of relevant energy can be provided by the QCD phase transition. In Ref.~\cite{Mupo:2025ner} it is proposed that, if the energy injected by the QCD phase transition through the Domain Wall network is enough to overcome the energy barrier for the non-minimal strings to decay, then this will efficiently drive the decomposition of the string network to a bunch of unit strings plus pure gauge strings.
Each unit axion string is attached to a single domain wall and the corresponding string-wall system subsequently collapses as in the standard $N_{\rm DW}=1$ scenario, while the pure gauge strings, having $w_a=0$, are not attached to the QCD domain walls. Rigorous numerical simulations are likely needed to actually demonstrate that the decomposition of the string network occurs given the energetic assumptions. For now, we are content to suppose the domain wall problem is solvable given the energetic condition needed for the string decomposition, as per~\cite{Mupo:2025ner}.

\subsection*{Estimating the domain wall constraint on our model parameter space}

Let's turn in this into an indicative constraint on the parameter space of our models (which should, of course, be taken with a large grain of salt, given the caveats just mentioned).
Consider a string with non-unit winding $|w_a|>1$ becomes attached to $|w_a|$ domain walls. The domain-wall tension is parametrically
\begin{equation}
    \sigma_{\rm DW}(T) \sim m_a(T)f_a^2.
\end{equation}
For the models we consider, the axion decay constant is $f_a = v_a/k$ where recall $\Z_k$ is the discrete symmetry left unbroken by the ABJ anomaly. Enforcing primitive solutions $k=1$, this becomes $f_a=v_a$, with $v_a$ given in~\eqref{eq:va}. For the class of models with partial lepton universality~\eqref{eq:r2r3}, this becomes $f_a \simeq v_1/(2|r|)$ in the $v_2 \gg v_1$ limit.

Over a characteristic length scale $L$ of the string network, the energy provided by the $|w_a|$ attached domain walls is of order $|w_a|\sigma_{\rm DW}L$. Taking $L\sim H^{-1}$, and assuming $v_2 \gg v_1$,\footnote{In the regime where $v_1$ is larger, the same formula but with $2 \leftrightarrow 1$ and $n\leftrightarrow m$ applies. When we come to plot the various constraints in Fig.~\ref{fig:post_inf}, we of course use the formula appropriate for each region. } the decomposition of the strings becomes energetically favourable when
\begin{equation}
|w_a|\,m_a(T_{\rm dec})f_a^2\,H^{-1}(T_{\rm dec})\gtrsim2\pi v_2^2\left[|w_a|(n_2^\ast)^2+|c|x_2^2-n^2\right]\, .
\label{eq:DW_decomposition_condition}
\end{equation}
The temperature $T_{\text{dec}}$ at which this condition is first satisfied therefore determines when the higher-winding strings can start decomposing. In order to solve the domain wall problem, this decay must occur before Big Bang Nucleosynthesis (BBN). We numerically solve for the two sides of~\eqref{eq:DW_decomposition_condition} being equal to find the cross-over temperature, then impose a constraint 
\begin{equation}
\label{eq:bbn_bound}
    T_{\text{dec}} \gg T_{\text{BBN}}
\end{equation}
on the parameter space of our models.
We take $T_{\text{BBN}} \sim 3$ MeV. Numerical simulations, needed to confirm a plausible estimate of $\mathcal{O}(T^{-1}_\textrm{dec})$ for the duration of this process, are beyond the scope of this paper. Also, these provide the weakest constraints on the parameter space of benchmark models discussed later.

\subsection{The stable relic problem}

The second main challenge in trying to realise post-inflationary axion models is to ensure there are no stable relics. 
In particular for our setups, the heavy quarks must have couplings to the SM that allow them to decay away rapidly before BBN time~\cite{DiLuzio:2016sbl,DiLuzio:2017pfr,Cheek:2023fht,DiLuzio:2024xnt}.  

Sufficiently fast decays are realised if the exotic fields can decay through operators with mass dimension $\leq 5$~\cite{DiLuzio:2016sbl,DiLuzio:2017pfr}.\footnote{Decays via operators with dimension larger than 5 can be acceptable in cosmological histories with an early matter domination period, see Refs.~\cite{Cheek:2023fht,DiLuzio:2024xnt}} We want to construct anomaly-free models for which this is the case for every VLQ representation required. We take all the VLQs to be in the down-type quark representation, $D \sim  (\bm{3}, \bm{1})_{-1/3}$. This choice, compared to using up-type quarks, turns out to be advantageous for postponing the hypercharge Landau pole as much as possible, see \S \ref{sec:Landau_Poles}. Decays will proceed via effective operators of the form
\begin{equation} \label{eq:O_alpha}
    \mathcal{O}_\alpha = \bar{D}_{L,\alpha}d_R \phi^{[p]}_1 \phi^{[q]}_2\, \textrm{ with }   |p|+|q| \leq  2, \qquad \mathrm{Dim}(\mathcal{O}_\alpha) = 3+ |p|+|q| \leq 5\, .
\end{equation}
Here negative values of $p$ or $q$ denote the corresponding conjugated scalar. 
For a given VLQ, a decay-induced effective operator involving the opposite VLQ chirality, which takes the form
$\mathcal{O}_\alpha^H = \overline q_L H D_{R,\alpha}\phi_1^{[r]}\phi_2^{[s]}$, can never appear at a lower dimension.\footnote{Because $L_\alpha-R_\alpha\in\{\pm X_1,\pm X_2\}$, switching $D_{R,\alpha}$ to $\bar{D}_{L,\alpha}$ can always be compensated by at most one additional insertion of $\phi_i$ or $\phi_i^\dagger$. Therefore, if the minimal operators involving $D_R$ and $D_L$ contain respectively $n_R$ and $n_L$ scalar insertions, $n_L\leq n_R+1$, and hence $\mathrm{Dim}(\mathcal O_\alpha)=3+n_L
\leq 4+n_R=\mathrm{Dim}(\mathcal O^H_\alpha)$. It is therefore sufficient to consider $\mathcal O_\alpha$ when identifying the lowest-dimensional VLQ decay operator.} These operators generically induce observable effects in the low-energy EFT of the SM, whose phenomenology has been studied in Refs.~\cite{Alonso-Alvarez:2023wig,Palavric:2026vej}.

These operators induce mixing between the VLQ and the SM quarks, after the $\phi_i$ fields acquire their VEVs. This opens decay channels involving SM bosons and quarks.\footnote{In general, decays of the VLQs into axions are also allowed, in the channels $D\to d\, a$, but these channels can be suppressed and we do not need to consider them.} 
For the slowest-decay case, {\em i.e.} where $\mathrm{Dim}(\mathcal{O}_\alpha)=5$, the
total width is approximately
\begin{equation}
\label{eq:decay_VLQs}
    \Gamma_D\simeq\Gamma_{D\to tW}+\Gamma_{D\to bZ}+\Gamma_{D\to bh}\simeq\frac{m_b^2}{8\pi v^2}\frac{M_D^3}{\Lambda^2}\simeq\frac{1}{8.5\times10^{-9}\,{\rm s}}\left(\frac{M_D}{10^9\,{\rm GeV}}\right)^3 ,
\end{equation}
where in the numerical estimate we have taken $v_1\sim v_2\sim M_D$. This is easily rapid enough to not disrupt BBN, regardless of the finer details that fix the size of the various order-1 couplings appearing here. So, if we can construct anomaly-free spectra of VLQs for which all species decay via operators of dimension $\leq 5$, then we have solved the stable relic problem. We build such models next.

\subsection{Anomaly-free spectra for post-inflationary benchmark models}
\label{sec:post_anomaly_cancellation}

Here we construct anomaly-free spectra of VLQs for viable post-inflationary models, for which the axion quality problem will be solved. Demanding there be no stable relics, as just described, and preserving axion quality (accounting also for PQ-violating operators with fermions), puts tight constraints on the chiral fermion spectrum. First, we describe how to solve the conditions coming from anomaly cancellation. 

We first note that, in contrast to the simplified (and relic-afflicted) setup considered in \S \ref{sec:2scalars}, our VLQs now have non-zero hypercharge to allow their decay, and so we have an additional anomaly cancellation condition to solve:
\begin{equation}
    \sum_\alpha N_\alpha\left(L^2_\alpha-R_\alpha^2\right) = 0 \, .
\end{equation}
coming from the mixed anomaly diagram involving one hypercharge boson and two $U(1)_X$ gauge bosons. This places a quadratic condition on the $X$ charges of the VLQs, in addition to the linear and cubic conditions.

Each BM model has fixed the two scalar charges $(X_1,X_2)=d_X(x_1,x_2)$, with $d_X$ either 9 or 18 depending on whether $d_{\text{PQ}}$ is odd or even (respectively), such that $\gcd(x_1,x_2)=1$. To find anomaly-free solutions, it is not enough to consider a single VLQ per $\phi_i$ as in \S \ref{sec:2scalars}; rather, we allow for $n$ types of VLQ representation labelled by index $\alpha = 1, \dots, n$ (with $n$ not fixed), with chiral $U(1)_X$ charges $(L_\alpha, R_\alpha)$ that are restricted such that
\begin{equation}
     X_\alpha := L_\alpha - R_\alpha \in \{\pm X_1, \pm X_2 \}\, ,
\end{equation}
which guarantees each VLQ has a dimension-4 Yukawa coupling to one of the scalars $\phi_i$ (or their conjugates).
Defining the sums of charges $S_\alpha:=L_\alpha+R_\alpha$ as before, and the multiplicity of that representation to be $N_\alpha$, the ACC system of equations is now
\begin{align}
    0 &= \sum_\alpha N_\alpha X_\alpha, \label{eq:X}\\
    0 &= \sum_\alpha N_\alpha X_\alpha S_\alpha, \label{eq:XX}\\
    0 &= \sum_\alpha N_\alpha X_\alpha (3 S_\alpha^2 + X_\alpha^2). \label{eq:XXX}
\end{align}
Now we additionally impose the condition to avoid stable relics, that is we require gauge invariance of the operator~\eqref{eq:O_alpha}. This translates to
\begin{equation}
    S_\alpha
    =
    -X_\alpha + 2m
    +2(p_\alpha X_1+q_\alpha X_2),
    \qquad
    |p_\alpha|+|q_\alpha|\leq 2.
\end{equation}
Defining
\begin{equation} \label{eq:r_alpha}
    r_\alpha:=p_\alpha x_1+q_\alpha x_2,
\end{equation}
the corresponding LH and RH charges are
\begin{equation} \label{eq:LandR}
    L_\alpha=m+d_Xr_\alpha,
    \qquad
    R_\alpha=m+d_X(r_\alpha-y_\alpha),
\end{equation}
where it was also convenient to define the rescaled charge differences $y_\alpha$ via
\begin{equation}
    X_\alpha = d_X y_\alpha,
\end{equation}
so that $y_\alpha \in \{\pm x_1,\pm x_2\}$. Finally, it is easy to verify that a common shift $S_\alpha \mapsto S_\alpha - 2m$ is a symmetry of the ACCs. For this reason, it is cleaner to trade $S_\alpha$ for a shifted (and rescaled) variable
\begin{equation}
    s_\alpha := 2r_\alpha - y_\alpha
\end{equation}
The variables $(y_\alpha, s_\alpha)$, which are re-scaled and shifted versions of the difference and sum of left and right charges, are useful variables to work with. 

Given $(y_\alpha, p_\alpha, q_\alpha)$ all live in small finite sets, it is already clear that there is only a small number of possible $L_\alpha$ and $R_\alpha$ charges that can be consistent with our conditions -- before we have even considered the constraints from anomaly cancellation and PQ quality (including operators involving VLQ bilinears). So, provided of course we can find {\em a} solution, there is a strong computational advantage coming from imposing our strict condition for avoiding stable relics.
Precisely, for each possible model {\em i.e.} each coprime pair $(x_1, x_2)$, there are only
\begin{equation*}
    4 \times 13 = 52    
\end{equation*}
candidate VLQ representations we can use, where $13$ is the number of integer pairs $(p,q)$ satisfying $|p|+|q| \leq 2$.
It might seem surprising that one should then find {\em any} viable solution, given the ACCs are a set of non-linear Diophantine equations. But sure enough, we are able to construct explicit BMs $(x_1, x_2, m)$ of interest for phenomenology. Our strategy from this point will be to (i) first filter for anomaly-free solutions, then (ii) filter for solutions where the leading PQ-breaking operator (allowing for arbitrary VLQ insertions) occurs at dimension $\geq d_{\rm PQ} = |x_1|+|x_2|$.

\subsubsection{Filtering through the ACCs: a moment problem}

In \S \ref{sec:2scalars} we showed that the PQ anomaly coefficient $\mathcal{A}_{\text{PQ}}$ ends up being equal to unity for the primitive solution to the linear ACC, that is for $(N_1, N_2) =\pm (x_2, -x_1)$ where recall $\gcd{(x_1,x_2)}=1$. We want to retain this feature here, for our Goldstone to be the QCD axion and to solve the domain wall problem. In this case, where we allow multiple species of VLQ coupled to each scalar $\phi_i$, the corresponding `primitive' condition is
\begin{equation} \label{eq:primitive_post}
    (\Delta_1,\Delta_2)=(x_2,-x_1), \qquad
    \text{where} \qquad
    \Delta_i := \sum_{y_\alpha=x_i}N_\alpha - \sum_{y_\alpha=-x_i}N_\alpha.
\end{equation}
This solves the linear ACC by construction, and enforces $\mathcal{A}_{\text{PQ}}=1$.
The two non-linear ACCs become, in our variables $(y_\alpha, s_\alpha)$,
\begin{align}
    0&=\sum_\alpha N_\alpha y_\alpha s_\alpha,
    \\
    0&=\sum_\alpha N_\alpha y_\alpha
    \left(
    3s_\alpha^2+y_\alpha^2
    \right)\, .
\end{align}
We emphasize that the parameter $m$ has dropped out of the analysis, thanks to the shift symmetry of the full ACC system that we noted above. We seek solutions to this pair of Diophantine equations, where the $r_\alpha$ variables run only over a finite range dictated by the stable relic condition $|p_\alpha|+|q_\alpha|\leq 2$.

Moreover, we would ideally like to find solutions with as few species of VLQ as possible. One motivation for this is the running of the QCD gauge coupling.  The total number of VLQs is
\begin{equation}
    N_{\text{VLQ}} := \sum_\alpha |N_\alpha|\, .
\end{equation}
The lower bound on the number of VLQs needed is, as in \S \ref{sec:2scalars},
\begin{equation}
    N_{\text{VLQ}}|_{\text{min}} = |\Delta_1|+|\Delta_2|=|x_1|+|x_2|,
\end{equation}
which recall equals $d_{\text{PQ}}$, the dimension of the first PQ-violating operator in the tree-level scalar potential. This is realised if there are no cancelling $\pm x_i$ pairs in the spectrum. That is, we have a total multiplicity of $|x_2|$ VLQs coupled to $\phi_1$ (or $\phi_1^\dagger$, but not both) and vice versa. More generally, the number of VLQs appearing in the spectrum can be expressed as
\begin{equation}
    N_{\rm VLQ} = d_{\rm PQ} + 2t, \qquad t \in \Z_{\geq 0} \, ,
\end{equation}
given extra VLQs (beyond the minimal number) necessarily appear in $\pm x_i$ pairs to preserve the solution to the linear anomaly constraint.

The remaining non-linear ACCs can be usefully recast as a moment problem, as follows.
First, it is helpful to thus divide our VLQ species into four sets, that we denote $\mathcal{S}_i^\pm$, depending on which of $\pm \phi_i^{(\dagger)}$ is used to write down its Yukawa coupling:
\begin{equation}
    \mathcal{S}_i^\pm:=\{\alpha\,|\,y_\alpha=\pm x_i\}\, .
\end{equation}
We can then define the (signed) moments:
\begin{equation}
    M_i^{(n)}
    := \sum_{\alpha\in \mathcal{S}_i^+} N_\alpha s_\alpha^n
    -  \sum_{\alpha\in \mathcal{S}_i^-} N_\alpha s_\alpha^n,
\end{equation}
The zeroth moments are simply the multiplicities $\Delta_i$ that we already introduced. Defining the weighted (and signed) means $\mu_i := M_i^{(1)}/M_i^{(0)}$,
the quadratic ACC is simply
\begin{equation} \label{eq:mean}
    \mu_1 = \mu_2 := \mu    
\end{equation}
That is, the $\mathcal{S}_1$ type VLQs and the $\mathcal{S}_2$ type VLQs have the same (signed and weighted) mean $\mu$.
Finally, the cubic ACC then becomes a condition on the second moment. Define the (signed and centred) second moments $V_i$ for each type of VLQ:
\begin{equation}
    V_i := \sum_{\alpha \in \mathcal{S}_i^+} N_\alpha (u_\alpha - \mu)^2 - \sum_{\alpha \in \mathcal{S}_i^-} N_\alpha (u_\alpha - \mu)^2 , \qquad i=1, 2.
\end{equation}
The cubic anomaly condition is then
\begin{equation}
    x_1 V_1 + x_2 V_2 = F(x_1, x_2), \qquad \text{where~~} F(x_1, x_2)=\frac{x_1 x_2}{3}(x_2^2-x_1^2)
\end{equation}
The computational challenge is thus reduced to computing the mean and variance for candidate $\mathcal{S}_1$ spectra and $\mathcal{S}_2$ spectra, matching their means, and then testing the condition on the second moments. Given the stable relic condition gives us a finite set of $r_\alpha$ to test (specified by integers $p_\alpha$ and $q_\alpha$ satisfying $|p_\alpha|+|q_\alpha| \leq 2$), viable solutions can be found with a small search starting from $N_{\rm VLQ}=d_{\rm PQ}$ species and and incrementing in steps of two. The final charges $(L_\alpha, R_\alpha)$ will end up appearing large, but that is mostly due to the scale factor of $d_X=9$ or $18$ appearing in the map~\eqref{eq:LandR}.

\subsubsection{Filtering through the PQ quality condition: a smallest distance problem} \label{sec:filterPQ}

Finally, bringing our construction full circle, we need to ensure there are no PQ-violating operators that we can build using the VLQs themselves that could spoil axion quality (assuming fermion legs can be closed in valid Feynman graphs to give loop corrections to the scalar potential), as discussed for the pre-inflationary models in \S \ref{sec:quality}. For our models with down-type VLQs, these potentially troublesome operators take the schematic form
\begin{align} \label{eq:PQV-fermion-Dtype}
    \mathcal{L} \supset \sum_k &\frac{1}{M_{\text{Pl}}^{3k+|s|+|t|-4}}\prod_{r=1}^k \left( \overline{D}_{L,\alpha_r} D_{R,\beta_r}\right) \phi_1^{[s]} \phi_2^{[t]}  \\
    &+\,\,\sum_l \frac{1}{M_{\text{Pl}}^{6l+|s|+|t|-4}} \prod_{p=1}^l  \left( \overline{D}_{X, \alpha_p} \gamma^\mu D_{X, \beta_p} \,\, \overline{D}_{Y, \gamma_p} \gamma_\mu D_{Y, \delta_p} \right)  \phi_1^{[u]} \phi_2^{[v]}  
    \label{eq:PQV_4F_type}
\end{align}
where the summations out front run over all possible operators, and we have omitted writing the Wilson coefficients explicitly, which we would expect to be order-1 in these units. In the second line, $X, Y \in \{L, R\}$. 

To ensure these operators do not spoil axion quality, we take a very conservative approach: we require {\em all} PQ-violating operators of the form~\eqref{eq:PQV-fermion-Dtype} and~\eqref{eq:PQV_4F_type} are banned to the same $d_{\text{PQ}} = |x_1|+|x_2|$ as the pure scalar operator -- irrespective of whether the fermion line can actually be closed to run into the scalar potential.\footnote{Other solutions are possible, which we explored. For instance, if closing the fermion loop requires non-renormalizable mixings, additional powers of $M_{\rm Pl}^{-1}$ enter the loop function in \cref{eq:grav_ferm_loop}, potentially restoring the required suppression. In practice, however, we generally find that such constructions also increase the dimension of the VLQ decay operators, although they can preserve the minimal multiplicity $N_{\rm VLQ}=d_{\rm PQ}$.}
Recall from \S \ref{sec:quality} that, if this is the case (and provided the fermion loops close), we expect such an operator to give a $k$-loop induced contribution to the scalar potential, suppressed by a factor
   $\left(n_c/16 \pi^2\right)^k (M_Q/v_i)^{3k}$
with respect to the pure-scalar operator already there in the tree-level potential, and so the fermion-induced contributions are subleading.

We first need the PQ charges of the VLQs, which we denote $(l_\alpha^{\rm PQ}, r_\alpha^{\rm PQ})$. We use the minimal normalisation for the PQ charges, {\em i.e.} such that the PQ  charges of the scalars $(\phi_1, \phi_2)$ are $(-x_2, x_1)$.
The PQ charges $(l_\alpha^{\rm PQ}, r_\alpha^{\rm PQ})$ are then fixed, independently of the parameter $m$, by the combination of the gauge-invariant Yukawa interaction ($\sim \overline{D}_{L,\alpha} \phi_i^{(\dagger)} D_{R,\alpha}$) and the gauge-invariant decay operator ($\sim \overline{D}_{L,\alpha} d_R \phi_1^p \phi_2^q$). The result is, for the LH component,
\begin{equation}
    l_\alpha^{\rm PQ} - q_f = -x_2 p_\alpha + x_1 q_\alpha\, .
\end{equation}
The charge of the RH component is the same but shifted by the PQ charge of the scalar needed to write down its Yukawa.
Note that the PQ charges are fixed by our Lagrangian only up to a common SM fermion PQ charge $q_f$, but this will drop out of any gauge-invariant operators we write down involving VLQ bilinears.

After doing the steps described above to find anomaly-free solutions that solve the stable relic problem, we wish to impose the refined condition on axion quality as a filter. This means removing possible charge assignments which allow operators of the form~\eqref{eq:PQV-fermion-Dtype} or~\eqref{eq:PQV_4F_type} up to dimension $d_{\rm PQ}$. We illustrate in more detail how this works for the chirality-flipping operators~\eqref{eq:PQV-fermion-Dtype}; a very similar approach is used for the chirality-preserving contractions~\eqref{eq:PQV_4F_type}.  

To simplify the task of scanning over all gauge-invariant operators of type~\eqref{eq:PQV-fermion-Dtype}, we observe that any fermion bilinear
\begin{equation}
    B_{\alpha\beta} := \overline{D}_{L,\alpha} D_{R,\beta}
\end{equation}
has the same gauge and PQ charges as a pure scalar operator
\begin{equation}
    O^\phi_{\alpha \beta} := \phi_1^{[c_1]} \phi_2^{[c_2]}
\end{equation}
with $c_1$ and $c_2$ uniquely determined thanks to the underlying charges being coprime. So, as far as charge counting goes, we can replace the fermionic operator with a scalar one, {\em viz.} $B_{\alpha\beta} \phi_1^{[s]} \phi_2^{[t]} \sim \phi_1^{[s+c_1]} \phi_2^{[t+c_2]}$, and similar for operators with more than one fermion bilinear insertion. Gauge invariance requires 
\begin{equation}
    (s+c_1, t+c_2) =n (-x_2, x_1) 
\end{equation}
using the fact that $\gcd{(x_1, x_2)}=1$. The $n=0$ case will preserve both $U(1)_X$ and PQ and so such operators do not threaten axion quality. But operators with non-zero $n$ will conserve $U(1)_X$ but necessarily violate PQ. This defines an integer lattice of permissible operators in the model that can spoil axion quality. 

As mentioned, we filter out models where {\em any} such operators occur with dimension $< d_{\rm PQ}$, which implies the leading PQ violation occurs from the tree-level pure scalar operator $\sim \phi_1^{[-x_2]} \phi_2^{[x_1]}$. To do this filtration, we divide the analysis into operators with fixed $k=1, 2, 3, \dots$ up to $k \leq \lfloor d_{\rm PQ}/3\rfloor$. Starting with $k=1$, and starting with some candidate VLQ spectrum that has passed the previous filtrations (including the ACCs), we compute all the independent values of $(c_1, c_2)$ that can be realised by the $n^2$ possible bilinears $B_{\alpha\beta}$ (recall $n$ is the number of different VLQ representations). Using the gauge invariance condition, this means we want to avoid there being any integer $n$ such that
\begin{equation}
    |x_2n+c_1| + |x_1n-c_2| < d_{\rm PQ} - 3
\end{equation}
for any $(c_1, c_2)$ furnished by our VLQ spectrum, where the shift by $3$ on the RHS is the mass dimension of $\bar{Q}Q$. One can verify this is true by plotting all the realised values of $(c_1, c_2)$ for the candidate spectrum, and showing that none lie within a diamond of side length $(d_{\rm PQ}-3)$ from the points $\pm (-x_2, x_1)$, which correspond to the first gauge-invariant (but PQ-breaking) points. 

\begin{figure}[tb]
    \centering
    \includegraphics[width=0.8\textwidth]{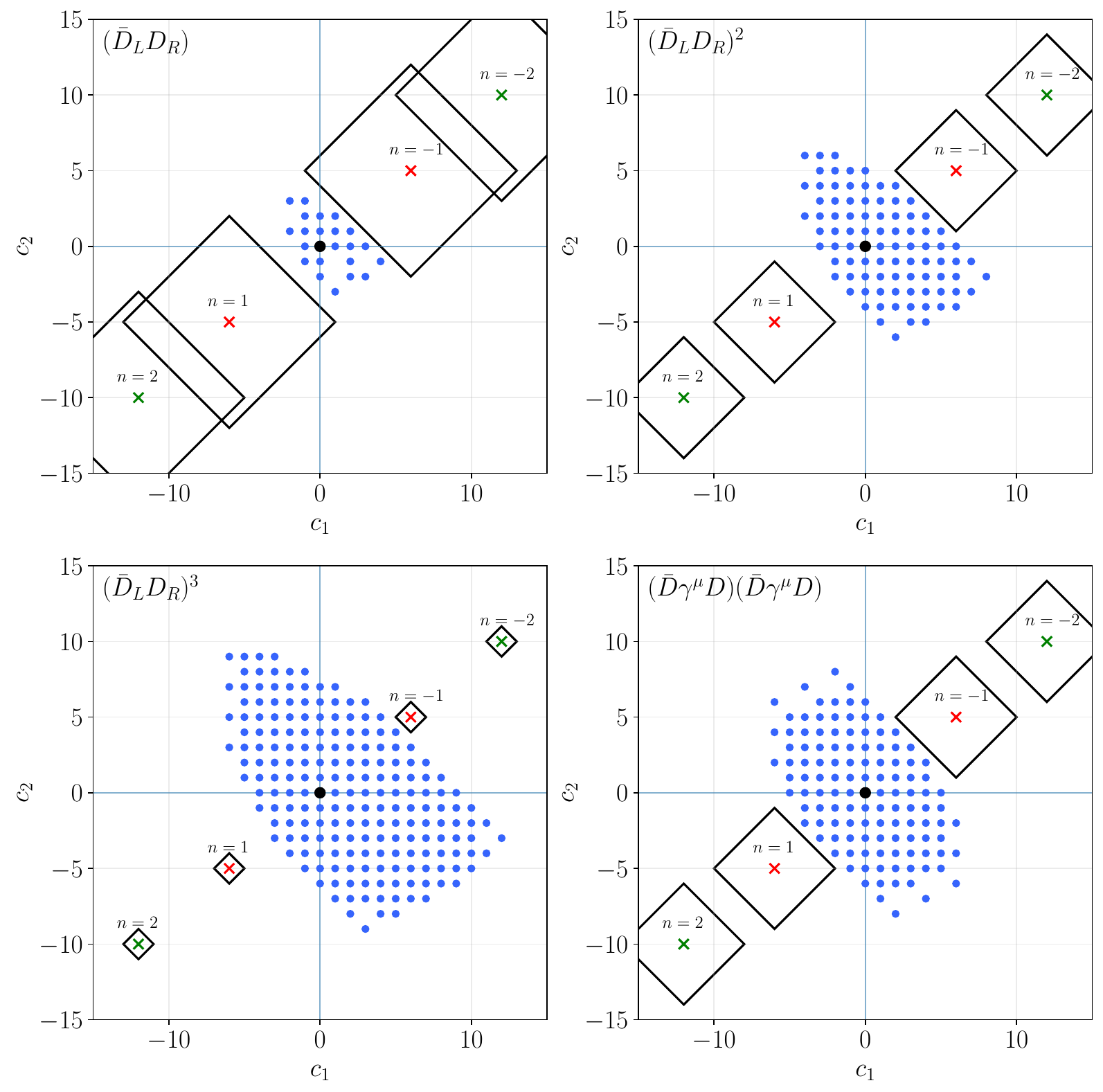}
    \caption{Graphical demonstration that there exist no gauge-invariant operators with fermion insertions, up to mass dimension $d_{\rm PQ}=11$ that violate PQ symmetry, for the benchmark model Post-1 (defined in Table~\ref{tab:BMd11}), which has scalar charges $(x_1,x_2)=(5,-6)$. 
    Blue dots denote PQ-breaking operators with the same charges as a VLQ bilinears, while crosses indicate points in the lattice corresponding corresponding to gauge-invariant operators. The square contours mark the regions in which replacing part of a scalar operator by an operator with fermions would yield a lower-dimensional, and therefore less suppressed, PQ-breaking operator. No blue dots enter any of these squares, demonstrating there are no PQ-breaking operators with dimension $< d_{\rm PQ}$.
    }
    \label{fig:quality-lattice}
\end{figure}

Since this may sound a little abstract, we illustrate this step for a benchmark model in Fig.~\ref{fig:quality-lattice}. An entirely analogous procedure is used to filter out models with PQ-violating operators involving same-chirality fermion contractions, of the form~\eqref{eq:PQV_4F_type}. We also illustrate in Fig.~\ref{fig:quality-lattice} how this type of operator is avoided for our benchmark models.  

\subsubsection{Defining the post-inflationary benchmark models}

With this preparation, we are ready to construct explicit VLQ spectra satisfying all our conditions to define two BM models with viable post-inflationary cosmology. 
We emphasize that each VLQ spectra remarkably satisfy all the following conditions:
\begin{enumerate}
    \item The SM $\times U(1)_X$ gauge symmetry is anomaly-free, while being completely chiral in the VLQs.
    \item All VLQs can decay to SM through dimension-5 operators, avoiding any problems associated with stable relics.
    \item There are no PQ-violating operators that can be built using the VLQs and scalars up to dimension $d_{\rm PQ}$.
\end{enumerate}
The phenomenology of these models (which does not depend explicitly on the VLQ completion, but only on the parameters $(x_1, x_2, m)$) is summarised in Fig.~\ref{fig:post_inf}.

\begin{table}[t]
\centering
\begin{tabular}{c c c c c c c c}
\toprule
BM & $d_{\rm PQ}$ & $(r_{S},r_{D})$ & $m$
& $(x_1,x_2)$ 
& Residual sym. & PQ breaking ops. \\
\midrule
Post-1 & 11 & $(8,-3)$   & $2$   & $(5,-6)$ 
& $\mathbb Z_{9}$
& $\phi_1^6 \phi_2^5$,~~ $(\overline{D}_{L,1} D_{R,6}) \phi_1^5 \phi_2^3$, ...  \\
Post-2 & 13 & $(4,3)$     & $10$  & $(7,6)$  
& $\mathbb Z_9$
& $(\phi_1^\ast)^6\phi_2^7$,~~ $(\overline{D}_{L,2} D_{R,1}) (\phi_1^\ast)^3 \phi_2^7$, ... \\

\bottomrule
\end{tabular}
\caption{Representative benchmark models with viable post-inflationary cosmology / phenomenology. The charges $(x_1,x_2)$ denote the primitive scalar charges; in both cases, the charges $X_i=9x_i$, hence the residual symmetry is precisely the proton-stabilising $\Z_9$. The last column gives examples of leading gauge-invariant operators that explicitly breaks the accidental PQ symmetry, up to Hermitian conjugation; the ... indicates there are more operators with fermions that have the same mass dimension. The spectra of VLQ representations, which are anomaly-free and give no stable relics, are shown in Tables~\ref{tab:BMd11} and~\ref{tab:BMd13} respectively. }
\label{tab:benchmark_models_post}
\end{table}

We define two benchmarks, one with the leading PQ violation occurring at $d_{\rm PQ}=11$, the other with $d_{\rm PQ}=13$. Both have viable phenomenology, as we discuss further in \S \ref{sec:post-inf-pheno} (see Fig.~\ref{fig:post_inf}). For both BMs, we will show that regions that solve the axion quality problem are compatible with axion searches, stellar cooling, and domain wall constraints. By construction, there are no stable relics. 
Moreover, there is a window of parameter space that can also accommodate leptogenesis. What's more, in the $d_{\rm PQ}=13$ case, the axion can also serve as all of the observed DM.

The spectra of VLQs constructed for these BMs are recorded in Tables~\ref{tab:BMd11} and~\ref{tab:BMd13}. Note that, even though the integers sampled through run over small ranges, the resulting charges end up large. This is simply a result of the multiplication by $d_X=9$ or 18 entering in the mapping back to $L_\alpha$ and $R_\alpha$.
The first BM model, with $d_{\text{PQ}} = 11$, has $(x_1, x_2, m)=(5, -6, 2)$. We find the minimal VLQ spectrum has $N_{\rm VLQ}=13$, so this is next-to-minimal (recalling $N_{\rm VLQ}|_{\rm min}=d_{\rm PQ}=11$ here). 
The second BM model, with $d_{\text{PQ}} = 13$, has $(x_1,x_2,m)=(7,6,10)$. For this benchmark, we find the minimal number of VLQs needed is 17. For both BMs, anomaly cancellation can be verified just by substituting the charges into the ACCs. The solution to the stable relic problem is trivially checked by inspecting the values of $(p_\alpha, q_\alpha)$. Recall the relations~\eqref{eq:r_alpha} and~\eqref{eq:LandR} to go between $y_\alpha, p_\alpha, q_\alpha$ and $L_\alpha, R_\alpha$.

\begin{table}[t]
\begin{center}
\begin{tabular}{c|c|c|c|c|c}
$\alpha$
& $N_\alpha$
& $y_\alpha$
& $(p_\alpha,q_\alpha)$
& $(L_\alpha,R_\alpha)$
& $(l^{\rm PQ}_\alpha-q_f,\;r^{\rm PQ}_\alpha-q_f)$
\\
\hline
1 & 1 & $-5$ & $(-1,1)$ & $(-97,-52)$  & $(-1,5)$ \\
2 & 2 & $-5$ & $(0,2)$  & $(-106,-61)$ & $(10,16)$ \\
3 & 2 & $-5$ & $(1,1)$  & $(-7,38)$    & $(11,17)$ \\
4 & 1 & $-5$ & $(2,0)$  & $(92,137)$   & $(12,18)$ \\
5 & 1 & $-6$ & $(1,0)$  & $(47,101)$   & $(6,1)$ \\
6 & 1 & $6$  & $(0,2)$  & $(-106,-160)$& $(10,15)$ \\
7 & 2 & $6$  & $(1,1)$  & $(-7,-61)$   & $(11,16)$ \\
8 & 3 & $6$  & $(2,0)$  & $(92,38)$    & $(12,17)$
\end{tabular}
    \caption{Anomaly-free VLQ spectrum for our benchmark model `Post-1', for which the first PQ-violating operators (allowing for arbitrary fermion insertions) occur at dimension 11, with $(x_1, x_2, m)=(5, -6, 2)$. All VLQs are in the same representation under $G_{\rm SM}$ as RH down-type quarks. The final column of the table records the charges under the global PQ symmetry of the VLQs. } 
    \label{tab:BMd11}
    \end{center}
\end{table}

\subsection*{Achieving full axion quality for the post-inflationary benchmark models }

It is perhaps useful to explain a little more how these BM models solve the axion quality problem as described in \S \ref{sec:filterPQ}, in other words how they avoid any gauge-invariant operators involving the VLQs up to dimension $d_{\rm PQ}$. 

Let us illustrate using the benchmark `Post-1' (Table~\ref{tab:BMd11}).
In this case there are $n=8$ types of VLQ representation, and we need to consider all operators with  either $k=1, 2$ or $3$ chirality-flipping bilinears (note $k=4$ would already be dimension-12 and so is permissible for this benchmark), as well as chirality-preserving four-fermion insertions. 

Consider first the chirality-flipping operators.
By considering each of the 64 possible bilinears $B_{\alpha \beta} = \overline{D}_{L,\alpha} D_{R,\beta}$, one finds there are 22 independent values of $(c_1,c_2)$ that can be realised.
To ban PQ-violating operators with $k=1$ fermion bilinear up to dimension $d_{\rm PQ} = 11$ means there exists no integer $n$ such that
\begin{equation} \label{eq:PQtest_Post1}
    |6n-c_1| + |5n-c_2| < 8\, ,
\end{equation}
for any of the $(c_1, c_2)$ values furnished by our VLQ spectrum. The top-left plot in Fig.~\ref{fig:quality-lattice} shows all the realised $(c_1, c_2)$ values as blue dots. Any solution to the inequality~\eqref{eq:PQtest_Post1} would lie inside one of the black diamonds, which are centred on the PQ-violating (but gauge-invariant) points $n(6, 5)$, $n \in \Z$, and have side length 7. The fact that none of the 22 blue points fall inside the black squares means there are no gauge-invariant operators that can be constructed with dimension $\leq 11$ that preserve $U(1)_X$ but violate PQ. Similar is shown to be true for operators with $k=2$ and $k=3$ fermion bilinears, as verified by the top right and bottom left plots respectively in Fig.~\ref{fig:quality-lattice}.

The bottom right plot in Fig.~\ref{fig:quality-lattice} demonstrates that PQ quality is also safe, up to $d_{\rm PQ}=11$, from operators involving four-fermion chirality-preserving insertions, $\mathcal{O}\sim (\bar D \gamma^\mu D)^2 \phi_1^{[u]}\phi_2^{[v]}$. In this case, the building blocks are the vector currents
\begin{equation}
    J_{L,\alpha\beta}^\mu := \overline{D}_{L,\alpha} \gamma^\mu D_{L,\beta}    
\end{equation}
and similarly $J_R^\mu$, and again each of these objects has the same $X$ and PQ charges as a pair of scalars; the (signed) number $(c_1, c_2)$ of scalars to which a given building block is charge equivalent is plotted as a blue dot. For the VLQ spectrum in the Post-1 benchmark, there are 19 independent charge vectors realised by the $J_{L,\alpha\beta}^\mu$, and 25 by the $J_{R,\alpha\beta}^\mu$. These are not all independent, however; only 31 independent charge vectors are realised, as displayed in the figure. The black squares centred on the $U(1)_X$-preserving (but PQ-violating) points here have side length $d_{\rm PQ}-1-6=4$, given the 4-fermion part has mass dimension-6, and so a blue dot lying inside a black square would indicate a PQ-violating operator at dimension lower than $d_{\rm PQ}$. There are no such offending points.  

\begin{figure}
    \centering
    \includegraphics[width=\linewidth]{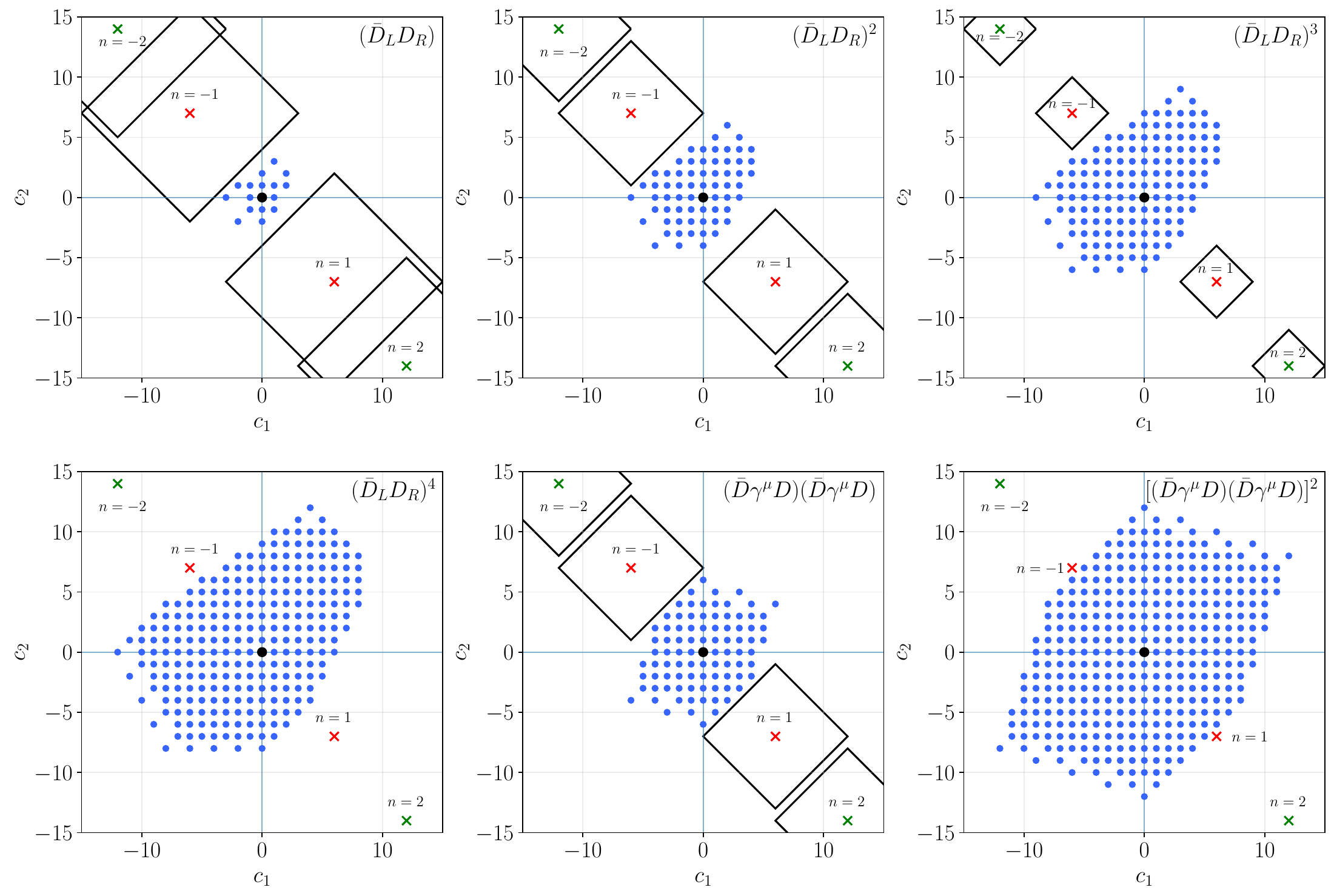}
    \caption{Graphical demonstration that there exist no gauge-invariant operators with fermion insertions, up to mass dimension $d_{\rm PQ}=13$ that violate PQ symmetry, for the benchmark model Post-2 (defined in Table~\ref{tab:BMd13}), which has scalar charges $(x_1,x_2)=(7,6)$. 
    Blue dots denote PQ-breaking operators with the same charges as a VLQ bilinears, while crosses indicate points in the lattice corresponding corresponding to gauge-invariant operators. The square contours mark the regions in which replacing part of a scalar operator by an operator with fermions would yield a lower-dimensional, and therefore less suppressed, PQ-breaking operator. No blue dots enter any of these squares, demonstrating there are no PQ-breaking operators with dimension $< d_{\rm PQ}$.}
    \label{fig:quality_post2}
\end{figure}

One can see from the plots in Fig.~\ref{fig:quality-lattice} that there exist blue points just `one step away' from the black diamond regions. This means there are operators involving fermions with mass dimension exactly equal to $d_{\rm PQ}$. The same occurs for both our benchmarks, Post-1 and Post-2, which reflects the fact that our solutions are minimal (in the sense of being those with the smallest $N_{\rm VLQ}$ that satisfy our criteria).
For benchmark Post-1, we have operators of each possible type hitting mass dimension $d_{\rm PQ}=11$. An example of each is given by
\begin{align}
    M_{\rm Pl}^7 \mathcal{L} \supset &(\overline{D}_{L,1} D_{R,6}) \phi_1^5 \phi_2^3 + (\overline{D}_{L,1} D_{R,6})^2 \phi_1^4\phi_2 +  (\overline{D}_{L,1} D_{R,6})^2 (\overline{D}_{L,1} D_{R,2}) \phi_1^2 
    \\ &+ (\overline{D}_{R,4} \gamma^\mu D_{R,5})(\overline{D}_{R,4} \gamma_\mu D_{R,5}) (\phi_1^\ast)^2 (\phi_2^\ast)^3
    + \dots
\end{align}
where we omit the Wilson coefficients for brevity.

The corresponding PQ-quality conditions for benchmark Post-2 require larger classes of operators to be scanned over, because the mass dimension we target is higher ($d_{\rm PQ}=13$ here). This means we need to consider operators with up to 8 fermions. The graphical demonstration of PQ quality is, for this benchmark, shown in Fig.~\ref{fig:quality_post2}. The story is very similar to that we explained for Post-1, only now there are six kinds of operator to check (and hence six sub-plots in the figure). 

For the benchmark Post-2, the leading PQ-violating operators with fermions occur at dimension-13. There are various kinds of operator structures that saturate this bound; operators with $k=1, 2,$ or 3 chirality flipping bilinears (but not, it turns out, 4), and operators with four or eight fermions contracted in a chirality preserving way.
An example of each kind of operator is given by:
\begin{align}
    M_{\rm Pl}^9 \mathcal{L}\,\, \supset &(\overline{D}_{L,2} D_{R,1}) (\phi_1^\ast)^3 \phi_2^7 + (\overline{D}_{L,2} D_{R,1})^2\phi_2^7 +  (\overline{D}_{L,5} D_{R,1})^3 \phi_2^4 
    \\ &+ (\overline{D}_{L,2} \gamma^\mu D_{L,3})(\overline{D}_{L,2} \gamma_\mu D_{L,3}) (\phi_1^\ast)^2 \phi_2^5 
    \\ &+ (\overline{D}_{L,2} \gamma^\mu D_{L,3})(\overline{D}_{L,2} \gamma_\mu D_{L,3})(\overline{D}_{L,2} \gamma^\mu D_{L,3})(\overline{D}_{L,5} \gamma_\mu D_{L,4}) \phi_2
    + \dots 
\end{align}
where we see the first appearance of a 8-fermion operator relevant to axion quality.

To close this discussion of PQ quality, it is important to revisit a comment we made previously, which is that one should consider whether operators like those we have written can in fact close via a loop diagram to give a shift in the scalar potential. 

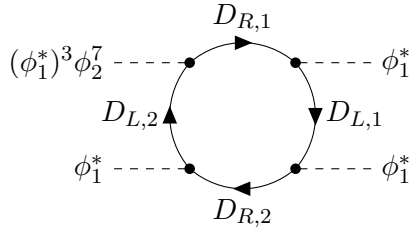
\begin{figure}[h]
\begin{center}
    \begin{tikzpicture}[baseline=-6.0ex]
    \begin{feynman}
        \vertex (v1) at (0,0);
        \vertex (v2) at (1.4,0);
        \vertex (v3) at (1.4,-1.4);
        \vertex (v4) at (0,-1.4);
        \vertex [left=1cm of v1] (a) {$(\phi_1^\ast)^3 \phi_2^7$};
        \vertex [right=1cm of v2] (c1) {$\phi_1^\ast$};
        \vertex [right=1cm of v3] (d) {$\phi_1^\ast$};
        \vertex [left=1cm of v4] (c) {$\phi_1^\ast$};
        \diagram*{
            (a) -- [scalar] (v1),
            (c1) -- [scalar] (v2),
            (d) -- [scalar] (v3),
            (c) -- [scalar] (v4),
            (v1) -- [fermion, bend left=40, edge label=$D_{R,1}$] (v2),
            (v2) -- [fermion, bend left=40, edge label=$D_{L,1}$] (v3),
            (v3) -- [fermion, bend left=40, edge label=$D_{R,2}$] (v4),
            (v4) -- [fermion, bend left=40, edge label=$D_{L,2}$] (v1)
        };
        \node at (v1) [circle,fill,inner sep=1.4pt]{};
        \node at (v2) [circle,fill,inner sep=1.4pt]{};
        \node at (v3) [circle,fill,inner sep=1.4pt]{};
        \node at (v4) [circle,fill,inner sep=1.4pt]{};
    \end{feynman}
    \end{tikzpicture}
    \caption{\label{fig:loop}
    Loop diagram allowing the dimension-13 operator $(\overline{D}_{L,2} D_{R,1}) (\phi_1^\ast)^3\phi_2^7$, inserted at the top left, to match into a dimension-13 pure-scalar operator $\sim (\phi_1^\ast)^6 \phi_2^7$ that violates PQ.
    }
\end{center}
\end{figure}

As an example, the operator $(\overline{D}_{L,2} D_{R,1}) (\phi_1^\ast)^3 \phi_2^7$ can be closed into a one-loop diagram, meaning it matches into a dimension-13 pure-scalar operator that violates PQ. The diagram is shown in Fig.~\ref{fig:loop}. 
Generically, if such operators with fermions are allowed, it is possible they can indeed match into the scalar potential in this way.
But the key point, as discussed, is that the loop suppression makes this a subleading PQ violating effect compared to the tree level $\phi_1^6 (\phi_2^\ast)^7$ contribution that we use to plot the axion quality `bound' in Fig.~\ref{fig:post_inf}.

\begin{table}[tb]
    \begin{center}
    \begin{tabular}{c|c|c|c|c|c}
    $\alpha$
    & $N_\alpha$
    & $y_\alpha$
    & $(p_\alpha,q_\alpha)$
    & $(L_\alpha,R_\alpha)$
    & $(l^{\rm PQ}_\alpha-q_f,\;r^{\rm PQ}_\alpha-q_f)$
    \\
    \hline
    1 & 2 & $7$  & $(-1,0)$ & $(-53,-116)$ & $(6,12)$ \\
    2 & 6 & $7$  & $(1,0)$  & $(73,10)$    & $(-6,0)$ \\
    3 & 1 & $-7$ & $(-1,1)$ & $(1,64)$     & $(13,7)$ \\
    4 & 1 & $-7$ & $(0,2)$  & $(118,181)$  & $(14,8)$ \\
    5 & 7 & $-6$ & $(0,-1)$ & $(-44,10)$   & $(-7,0)$
    \end{tabular}
    \caption{Anomaly-free VLQ spectrum for our benchmark model `Post-2', for which the first PQ-violating operators (allowing for arbitrary fermion insertions) occur at dimension 13, with $(x_1,x_2,m)=(7,6,10)$. All VLQs are in the same representation under $G_{\rm SM}$ as RH down-type quarks. This solution has $N_{\rm VLQ}=17$. }
    \label{tab:BMd13}
    \end{center}
\end{table}

\subsection{Landau poles} \label{sec:Landau_Poles}

The benchmark models we have constructed for post-inflationary axion models feature a large number of VLQs. As we saw, the number of VLQs can be expressed as $N_{\rm VLQ}=d_{\rm PQ}+2t$, $t \in \Z_{\geq 0}$, and axion quality requires $d_{\rm PQ} \geq 10$, and typically larger (given we are interested in regions that can also explain leptogenesis, which pushes us to larger VEVs than is strictly needed for the axion only). 

This large number of vector-like quarks can significantly modify the running of the SM gauge couplings. At one loop,
\begin{equation}
    \frac{{\rm d} g_i}{{\rm d}\log\mu}
    =
    \frac{b_i}{16\pi^2}g_i^3\, ,
\end{equation}
such that, above the VLQ threshold $M_Q$, sufficiently large multiplicities can generate a Landau pole below the Planck scale. We wish to avoid this loss of control at high scales, and require all gauge couplings to remain perturbative up to $M_{\rm Pl}$.

\begin{figure}
    \centering
    \includegraphics[width=0.75\linewidth]{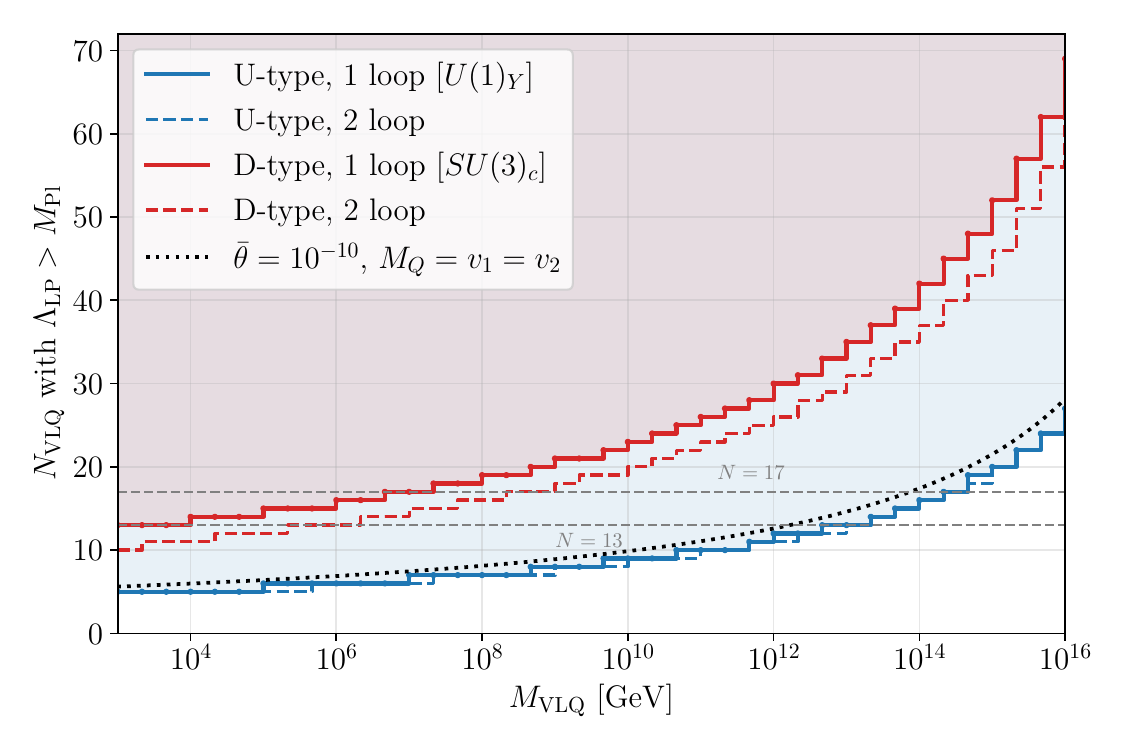}
    \caption{Maximum number of vector-like quarks compatible with perturbativity up to the Planck scale, as a function of their common mass $M_{\rm VLQ}$. Solid and dashed lines show the one- and two-loop results, respectively. For $U$-type VLQs the strongest constraint arises from the running of $U(1)_Y$, while for $D$-type VLQs it is set by $SU(3)_c$. The dotted black line indicates, for comparison, the minimum $d_{\rm PQ}=N_Q$ required by the axion-quality condition for the illustrative choice $v_1=v_2=M_Q$; so we see that models using $U$-type VLQs are disfavoured by axion quality, while those using $D$-types are compatible while retaining perturbativity.}
    \label{fig:landau_poles}
\end{figure}
 
This translates to a constraint on $N_{\rm VLQ}$. The maximal allowed multiplicity is shown in \cref{fig:landau_poles}, as a function of the VLQ mass (which we assume is the same for all VLQ species, for simplicity). We show results for VLQs transforming either all as RH up-type quarks (under the SM gauge group), which we refer to as $U$-type VLQs, and similarly for $D$-types. For $U$-type VLQs, their larger hypercharge makes the $U(1)_Y$ coupling diverge faster, and avoiding the hypercharge Landau pole provides the limiting constraint (blue lines in Fig.~\ref{fig:landau_poles}). For masses of order $v_i \sim 10^{11}$ GeV, this limits $N_{\rm VLQ} \lesssim 10$ or so, which is typically less than we need for our solutions to the axion quality problem. To highlight this tension, we show on Fig.~\ref{fig:landau_poles} the minimum $d_{\rm PQ}$ required by the axion-quality condition for the illustrative choice $v_1=v_2=M_Q$ (dotted black line). 

Conversely, for $D$-type VLQs the hypercharge contribution is substantially smaller, and the strongest restriction on $N_{\rm VLQ}$ arises from avoiding a Landau pole for $SU(3)_c$.\footnote{A stronger condition would be obtained by requiring the QCD one-loop beta function remains negative as in the SM, which would preserve asymptotic freedom. That would require $N_{\rm VLQ} \leq 10$, irrespective of $M_Q$. The only benchmark model considered in this paper which preserves this feature is `Pre-1', which has $d_{\rm PQ}=10$. Interestingly, it has viable phenomenology consistent with axion quality and all observational constraints, as shown in Fig.~\ref{fig:pre_inf_models}. However, that benchmark cannot simultaneously account for dark matter. } The two-loop running leads to somewhat stronger bounds, while preserving the same qualitative behaviour. This time, for VLQ masses of order $v_i \sim 10^{11}$ GeV, we can retain perturbativity as long as $N_{\rm VLQ} \lesssim 20$ or so, as satisfied by all the benchmark models we consider in this paper. Most dangerous is the benchmark `Post-2' (Table~\ref{tab:BMd13}), which has $N_{\rm VLQ}=17$.

The large number of fields charged under $U(1)_X$ could, in principle, force the gauge coupling $g_X$ to be small if perturbativity is required up to the Planck scale. Importantly, however, this does not necessarily imply that $X$ has to be light compared with the RH neutrinos. At one loop, the absence of a Landau pole below $M_{\rm Pl}$ implies schematically
\begin{equation}
    g_X^2 \lesssim
    \frac{8\pi^2}
    {b_X\log(M_{\rm Pl}/\mu)}\, ,
\end{equation}
where $b_X$ contains contributions of all fields active in the running. Here we have taken $\mu$ as a common threshold for simplicity. In the full numerical analysis, the different states are instead matched at their corresponding mass thresholds and evolved piecewise up to $M_{\rm Pl}$. With the mass of the gauge boson $M_{X} =g_X\sqrt{X_1^2v_1^2+X_2^2v_2^2}$, and combining with the previous expression, we obtain the largest gauge-boson mass compatible with perturbativity,
\begin{equation}
    M_{X}^{\rm max}
    \sim
    \sqrt{\frac{8\pi^2}
    {b_X\log(M_{\rm Pl}/\mu)}}
    \sqrt{X_1^2v_1^2+X_2^2v_2^2}\, .
\end{equation}
The compatibility with the heavy-$X$ regime relevant for leptogenesis can be understood analytically in the hierarchical limit $v_2\gg v_1$. In this regime,
\begin{equation}
    \frac{M_X^{\rm max}}{M_{N_1}}
    \sim
    |X_2|
    \sqrt{
        \frac{8\pi^2}
        {b_X\log(M_{\rm Pl}/\mu)}
    }
    \left(\frac{v_2}{v_1}\right)^2 ,
\end{equation}
and, even if the large charged matter content reduces the maximal value of $g_X$, a moderate hierarchy between the two symmetry-breaking scales rapidly compensates for this effect. Therefore, these models can remain perturbative below the Planck scale while successfully realising leptogenesis. 

Furthermore, we have checked that we can always find a large enough $g_X$ coupling, consistent with perturbativity to the Planck scale, such that the EFT validity of of the leptogenesis constraint in \cref{eq:leptogen} is satisfied.

\subsection{Parameter space of post-inflationary benchmark models} \label{sec:post-inf-pheno}

\begin{figure}[tb]
    \centering
    \includegraphics[width=\linewidth]{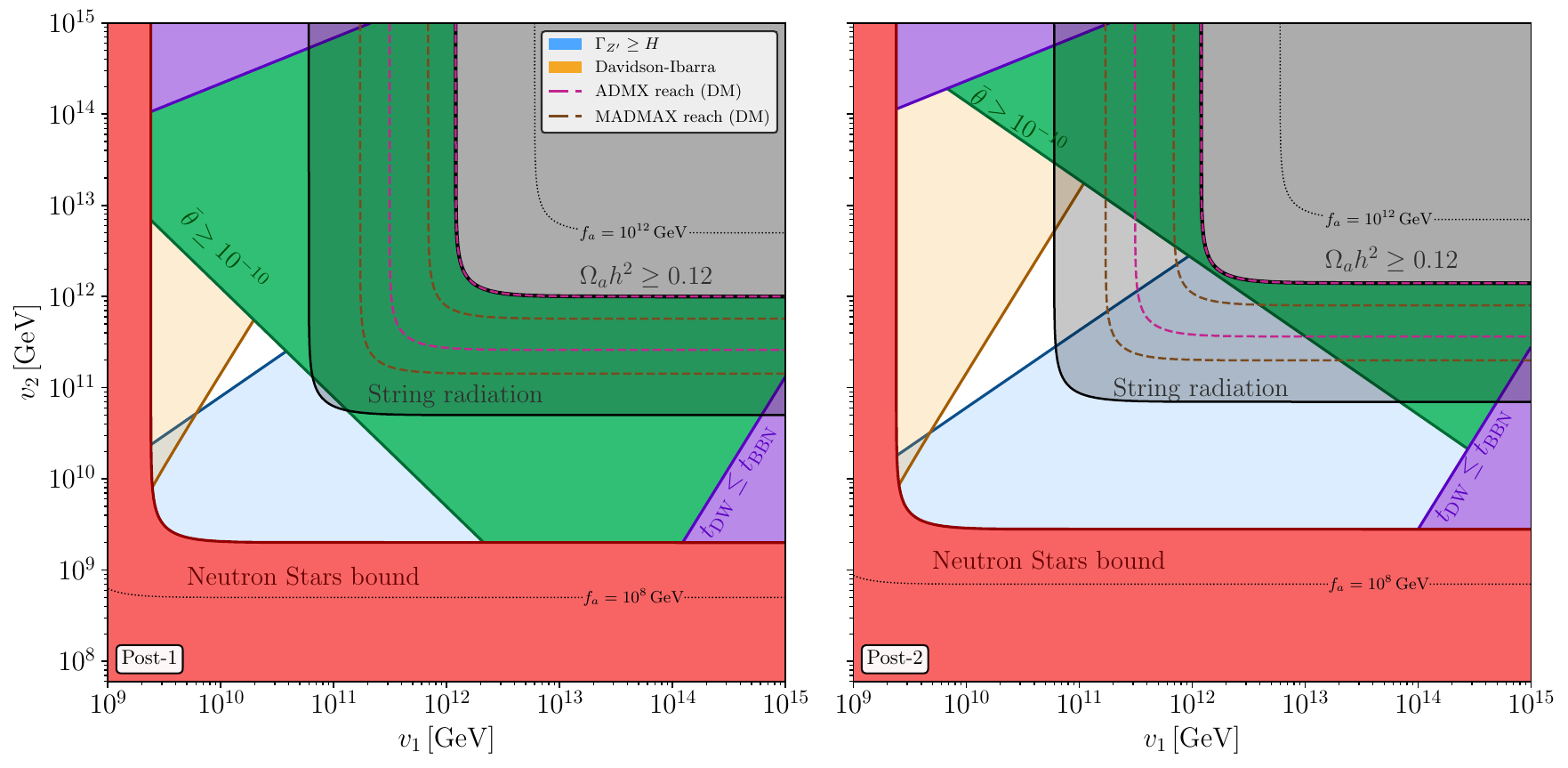}
    \caption{
    Phenomenology of the benchmark models labelled `Post-1' and `Post-2' (defined in Table~\ref{tab:benchmark_models_post}), that have viable phenomenology with the PQ transition occurring post inflation, shown in the $(v_1,\,v_2)$ plane. The PQ-quality bound from $\bar\theta$ is shown in green; Post-1 (Post-2) have the first PQ-violating operators occuring at dimension 11 (13) respectively, hence Post-2 has a larger region compatible with axion quality. The requirement that the string--domain-wall network decays before BBN (\S \ref{sec:DomainWalls}), which is a new feature relevant only for these post-inflationary models, is shown in purple. Dashed contours indicate the projected reach of IAXO, ADMX and MADMAX, while the red region is excluded by neutron-star cooling. The black contour marks $f_a=2\times10^{11}\,\mathrm{GeV}$, while the grey regions indicate the axion dark-matter abundance from misalignment and string radiation.
    Finally, the regions that are incompatible with leptogenesis are shown in blue and orange with lower opacity (because these are not necessary constraints that must be satisfied); in particular, the Davidson--Ibarra bound is in orange and the decoupling of gauge interactions in blue. }
    \label{fig:post_inf}
\end{figure}

Finally, we put everything together and summarise the constraints on the two post-inflationary benchmark models we have constructed, summarised in Table~\ref{tab:benchmark_models_post}, in Fig.~\ref{fig:post_inf}.

These benchmark models feature a number of $D$-type VLQs in their completion, from which we can deduce the axion coupling to photons. The coupling of the axion to photons is again particularly simple in this class of models. Indeed, all the heavy quarks introduced above transform in the same SM representation, $D\sim(\bm{3},\bm{1})_{-1/3}$, giving 
\begin{equation}
    \mathcal{L}_{\rm axion} = \frac{E}{32\pi^2} \frac{a}{f_a} F \widetilde{F} \,  \, \to \, \,  E= 6 Y_D^2 \frac{d_X}{ v_X^2}\left(\Delta_1 X_2 v_2^2 - \Delta_2 X_1 v_1^2\right) = \frac{2}{3}\frac{\Delta_1}{x_2} =\frac{2}{3} \, .
\end{equation}
The final equality used the fact that our benchmarks are built from primitive solutions, in the sense of~\eqref{eq:primitive_post}.
The fact that both benchmarks require additional pairs of VLQs with charge $\pm x_i$ gives cancelling contributions to the axion photon coupling. 

The bounds from leptogenesis (blue and orange, light transparency) and axion quality (green) depend only on the VEVs $v_1$ and $v_2$ as before, and we here plot them for the new values of $(x_1, x_2)$.
Compared to the phenomenology of the pre-inflationary models shown in Fig.~\ref{fig:pre_inf_models}, the constraints from isocurvature and from avoiding thermal restoration of the PQ symmetry are no longer present. However, we here have an additional constraint from requiring the sufficiently fast decay of the domain wall plus string network. Imposing the condition that strings decompose before BBN, as per \cref{eq:DW_decomposition_condition,eq:bbn_bound}, we obtain the purple regions in \cref{fig:post_inf}. Regarding the stable relic problem, we reiterate that our benchmarks were constructed hard-coding the requirement that all VLQs can decay to the SM through dimension-5 operators (see~\cref{eq:decay_VLQs}), so we can safely assume that the VLQs decay before BBN. 

As discussed in \cref{sec:DM}, the cosmological history is also different, and since the value of $\langle\theta_i^2\rangle$ is fixed, any value over $f_a\gtrsim 2\times10^{11}$ will overproduce DM. Furthermore, taking the range of cosmic string radiation for the minimal QCD axion~\cite{Benabou:2024msj, Saikawa:2024bta}, we see that the typical values of $f_a$ compatible with DM extend the parameter space to the lower-left regions of \cref{fig:post_inf}. Our benchmark scenario Post-2 has a reasonably wide region compatible with the DM range estimated from cosmic strings, that can also explain leptogenesis. We also show the region that is testable by future ADMX and MADMAX searches~\cite{Stern:2016bbw,Beurthey:2020yuq}.

In summary,~\cref{fig:post_inf} demonstrates that the axion models we constructed in this paper enjoy a consistent post-inflationary cosmological history, and at the same time can account for the origin both of ordinary baryonic matter (and its asymmetry with respect to anti-matter) and of dark matter, while remaining compatible with stringent PQ-quality requirements. 
Finally, we recall the starting point of our construction, which is that $U(1)_X$-breaking predicts exact proton stability and gives an excellent fit to neutrino mass and mixing data.

\section{Conclusions}
\label{sec:concl}

We speculate that proton stability and axion quality might share the same origin. Starting from the anomaly-free $U(1)_X$ construction of Ref.~\cite{Davighi:2022qgb}, spontaneous breaking leaves the exact $\mathbb Z_9$ selection rule $\Delta B=0\pmod 3$ predicting proton stability. 
If that breaking is triggered by a pair of scalar fields $\phi_{1,2}$ with certain charges, the same $U(1)_X$-breaking can generate the Majorana masses of the right-handed neutrinos, while leaving one physical Goldstone mode. Such a simple setup is remarkably rich in phenomenology: it can simultaneously accommodate neutrino masses and mixings, thermal leptogenesis, and the dark matter relic abundance of the Goldstone~\cite{Greljo:2025suh}, addressing many of the open empirical questions that require physics beyond the SM. 

By adding heavy coloured fermions that are vector-like under the SM but chiral under $U(1)_X$, this Goldstone becomes a QCD axion. In this paper, we show how the same gauge-charge assignment that provides the exact baryon-number selection rule can control the leading explicit breaking of the accidental Peccei--Quinn symmetry, to deliver axion models with the quality problem solved. The leading purely scalar PQ-breaking operator has dimension $d_{\rm PQ}=|x_1|+|x_2|$, where $x_i$ are the primitive charges of $\phi_i$, see \S\ref{sec:2scalars}. Importantly, this scalar counting is not by itself sufficient: operators containing heavy quarks may provide a lower-dimensional route to PQ violation once the fermions are matched onto the scalar potential. We therefore imposed the more conservative requirement that all PQ-charged operators be absent below mass dimension $d_{\rm PQ}$, including those with fermions, and find explicit constructions that pass this full operator test for PQ quality. 

Models with larger $d_{\rm PQ}$ (and thus better axion quality) necessarily require more heavy quark fields in their anomaly-free completion. We thus encounter an apparent tension between axion quality and maintaining perturbativity of the SM gauge couplings. We find only one viable benchmark model, with $d_{\rm PQ}=10$, that solves the quality problem while retaining asymptotic freedom of the QCD gauge coupling. All our benchmarks, nevertheless, remain under perturbative control up to the Planck scale. 

We demonstrate the flexibility of this model-building framework by constructing models consistent with PQ-breaking transitions that occur either before or after inflation. We begin by considering the pre-inflationary breaking scenario, since this provides the cleanest cosmology. Inflation removes the string network and any primordial heavy-quark relics, provided that reheating does not restore the PQ symmetry. These models are easier to construct than their post-inflationary siblings, in part because we can entertain heavy quarks that are neutral under electroweak symmetry. The remaining requirements come from axion isocurvature, from requiring a sufficiently low reheating temperature to preserve the broken phase, and from demanding a sufficiently high reheating temperature to populate the lightest right-handed neutrino for thermal leptogenesis. These conditions leave only restricted windows in the $(v_1,v_2)$ plane. But strikingly, as shown in \cref{fig:pre_inf_models}, we find benchmark models with parameter regions in which exact proton stability, realistic neutrino masses, high axion quality, thermal leptogenesis, and fitting the full dark-matter abundance (via the axion) can coexist. 

In \S \ref{sec:post_inflationary} we then build versions of the model that have viable post-inflationary cosmology, while retaining excellent axion quality (amongst our other {\em desiderata}). The post-inflationary scenario is a challenge to realise, requiring two additional cosmological conditions.  First, the thermally produced heavy quarks must decay before BBN. This significantly restricts the possible heavy-quark representations available to us (for fixed $x_i$), while at the same time introducing a mixed anomaly constraint that puts a quadratic condition on their charges. Taking into account also the stringent requirements from axion quality on banning PQ-violating operators involving heavy-quark bilinears, it is an intricate challenge to build viable UV completions in the post-inflationary regime, but one we overcome in \S \ref{sec:post_anomaly_cancellation}. 

The second cosmological condition for post-inflationary axion models is that the string--domain-wall network must disappear (\S \ref{sec:DomainWalls}). Although the primitive anomaly coefficient $\mathcal A_{\rm PQ}=1$ implies a single inequivalent QCD vacuum (enjoyed by all our models), individual strings in two-scalar theories like ours can carry non-unit axion winding and therefore have several domain walls attached to them. Such strings are, nevertheless, topologically permitted to split into unit-winding axion strings and pure-gauge strings~\cite{Mupo:2025ner}. We restrict to regions of parameter space in which the energy released by the QCD walls is sufficient to make this splitting energetically favourable before BBN.\footnote{We acknowledge that numerical simulations are, however, needed to demonstrate the conditions under which the decomposition of the string-wall network will indeed proceed efficiently; the viability of our post-inflationary models should (as for many other models) be understood subject to this caveat.} 

The upshot is, as shown in \cref{fig:post_inf}, that our post-inflationary benchmark models are also compatible with axion dark matter, the necessary conditions for thermal leptogenesis, and high axion quality.

At experimentally accessible energies, both constructions reduce to predictive KSVZ-like axion models. In the pre-inflationary completion, the heavy quarks are electroweak singlets with vanishing hypercharge, giving $E/N=0$, while the decay-safe down-type completion relevant for the post-inflationary scenario predicts $E/N=2/3$. The axion--photon coupling is therefore fixed in each case. Neutron-star cooling and future helioscopes such as IAXO probe these models independently of the axion relic abundance, whereas ADMX and MADMAX directly test the regions in which the axion constitutes the dark matter. The two-stage symmetry breaking transition triggered by our pair of scalars in the post-inflationary scenario can even provide an interesting target for gravitational wave experiments, with perhaps a distinct signal~\cite{Mupo:2025ner} noting that leptogenesis favours some hierarchy between $v_1$ and $v_2$. The models we build here have sharp predictions for ever-improving measurements of the neutrino mass and mixing parameters~\cite{Greljo:2025suh}. These ongoing experimental efforts on many fronts give hope that these models can be tested in the near future.

\section*{Acknowledgements}

The work of JD was supported by the Science and Technology Facilities Council (STFC) through an Ernest Rutherford Fellowship, under grant UKRI/ST/C002428/1, and was partially supported by the STFC HEP consolidated grant ST/X000664/1.

\bibliographystyle{utphys}
\bibliography{bibliography.bib}

\end{document}